\documentclass[journal,draftcls,onecolumn,12pt,twoside]{IEEEtran}
\IEEEoverridecommandlockouts
\usepackage{subfigure}

\usepackage{graphicx,amsmath,amssymb,amsfonts,cite}
\usepackage{algorithm}
\usepackage{setspace}
\usepackage{booktabs}
\usepackage[utf8]{inputenc}
\usepackage{algpseudocode}
\usepackage{bbold}
\makeatletter
\def\algbackskip{\hskip-\ALG@thistlm}
\makeatother
\usepackage{rotating}
\usepackage{makecell, multirow}
\renewcommand\theadfont{\normalsize}

\usepackage[skip=1ex]{caption}

\usepackage{float}
\usepackage{amsthm}
\usepackage[english]{babel}
\usepackage{ulem}
\usepackage{bm}
\usepackage{array}
\usepackage{color}
\usepackage{caption}
\ifCLASSINFOpdf
\else
\fi
\begin{document}
%
% paper title
% can use linebreaks \\ within to get better formatting as desired
\title{Machine Learning Enabled Preamble Collision Resolution in Distributed Massive MIMO
}

\author{Jie~Ding,
Daiming~Qu, Pei Liu, and Jinho Choi,~\IEEEmembership{Senior Member,~IEEE}
	% <-this % stops a space
	% \vspace{-4ex}
\thanks{Jie Ding and Jinho Choi are with the School of Information Technology, Deakin University, Geelong, VIC 3220, Australia.
}
\thanks{Daiming Qu is with the
School of Electronic Information
and Communications, Huazhong University of Science and Technology,
Wuhan, 430074, China.}
\thanks{Pei Liu is with the School of Information Engineering, Wuhan University of Technology, Wuhan, 430070, China.}
\thanks{This work was supported by the National Natural Science Foundation of China (NSFC) under Grant 61701186.}
}
% author names and affiliations
% use a multiple column layout for up to three different
% affiliations
% conference papers do not typically use \thanks and this command
% is locked out in conference mode. If really needed, such as for
% the acknowledgment of grants, issue a \IEEEoverridecommandlockouts
% after \documentclass

% for over three affiliations, or if they all won't fit within the width
% of the page, use this alternative format:
%%
%\author{\IEEEauthorblockN{Michael Shell\IEEEauthorrefmark{1},
%Homer Simpson\IEEEauthorrefmark{2},
%James Kirk\IEEEauthorrefmark{3},
%Montgomery Scott\IEEEauthorrefmark{3} and
%Eldon Tyrell\IEEEauthorrefmark{4}}
%\IEEEauthorblockA{\IEEEauthorrefmark{1}School of Electrical and Computer Engineering\\
%Georgia Institute of Technology,
%Atlanta, Georgia 30332--0250\\ Email: see http://www.michaelshell.org/contact.html}
%\IEEEauthorblockA{\IEEEauthorrefmark{2}Twentieth Century Fox, Springfield, USA\\
%Email: homer@thesimpsons.com}
%\IEEEauthorblockA{\IEEEauthorrefmark{3}Starfleet Academy, San Francisco, California 96678-2391\\
%Telephone: (800) 555--1212, Fax: (888) 555--1212}
%\IEEEauthorblockA{\IEEEauthorrefmark{4}Tyrell Inc., 123 Replicant Street, Los Angeles, California 90210--4321}}

%\thanks{Copyright (c) 2012 IEEE. Personal use of this material is permitted. However, permission to use this material for any other purposes must be obtained from the IEEE by sending a request to pubs-permissions@ieee.org.}}
% use for special paper notices
%\IEEEspecialpapernotice{(Invited Paper)}

\maketitle

\begin{abstract}
Preamble collision is a bottleneck that impairs the performance of random access (RA) user equipment (UE) in grant-free RA (GFRA).
In this paper, by leveraging distributed massive multiple input multiple output (mMIMO) together with machine learning, a novel machine learning based framework solution is proposed to address the preamble collision problem in GFRA. The key idea is to identify and employ the neighboring access points (APs) of a collided RA UE for its data decoding rather than all the APs, so that the mutual interference among collided RA UEs can be effectively mitigated. To this end,
we first design a tailored deep neural network (DNN) to enable the preamble multiplicity estimation in GFRA, where an energy detection (ED) method is also proposed for performance comparison. With the estimated preamble multiplicity, we then propose a $K$-means AP clustering algorithm to cluster the neighboring APs of collided RA UEs and organize each AP cluster to decode the received data individually. Simulation results show that a decent performance of preamble multiplicity estimation in terms of accuracy and reliability can be achieved by the proposed DNN, and confirm that the proposed schemes are effective in preamble collision resolution in GFRA, which are able to achieve a near-optimal performance in terms of uplink achievable rate per collided RA UE, and offer significant performance improvement over traditional schemes.
\end{abstract}
% IEEEtran.cls defaults to using nonbold math in the Abstract.
% This preserves the distinction between vectors and scalars. However,
% if the conference you are submitting to favors bold math in the abstract,
% then you can use LaTeX's standard command \boldmath at the very start
% of the abstract to achieve this. Many IEEE journals/conferences frown on
% math in the abstract anyway.
\begin{IEEEkeywords}
Preamble collision resolution, grant-free random access, deep learning, distributed massive MIMO, clustering.
\end{IEEEkeywords}
% no keywords
%\begin{IEEEkeywords}
%Success probability, grant-free, random access, massive MIMO, M2M.
%\end{IEEEkeywords}

% For peer review papers, you can put extra information on the cover
% page as needed:
% \ifCLASSOPTIONpeerreview
% \begin{center} \bfseries EDICS Category: 3-BBND \end{center}
% \fi
%
% For peerreview papers, this IEEEtran command inserts a page break and
% creates the second title. It will be ignored for other modes.
\IEEEpeerreviewmaketitle

\section{Introduction}
The Fifth Generation (5G) and future wireless communication focus on three major communication categories: enhanced mobile broadband (eMBB), ultra-reliable low-latency communication (URLLC), and massive machine-type communication (MTC) (mMTC) \cite{1}. Among them, mMTC has been regarded as an essential communication paradigm for a wide range of applications including healthcare, smart home, smart agriculture, and logistics and tracking \cite{2}. Since mMTC usually features with massive uplink access and limited packet size in nature,
it imposes new requirements and challenges in terms of random access (RA) design \cite{3}.

In Long Term Evolution (LTE) systems, a typical grant-based RA procedure is used to provide reliable access for human-type communication (HTC) \cite{3_0}. Since the grant-based RA requires handshaking to issue exclusive channel reservation for each RA user equipment (UE), it is unable to support massive access due to limited channel-resource utilization efficiency and also results in high signalling overhead to mMTC RA UEs. In the light of this, grant-free RA (GFRA) procedure has been recently actively studied in MTC for low signalling overhead and latency \cite{4_0,4_1,4,5}. In GFRA, the request-grant handshaking steps in grant-based RA for channel reservations are skipped, which allows RA UEs to access the network without grant acquisition once they have data to send \cite{4}. As a result, the signalling overhead is reduced \cite{5}. In \cite{4_0,4_1}, GFRA schemes by utilizing spreading techniques are proposed to support MTC. Nevertheless, spreading results in a bandwidth
expansion, which is inefficient in terms of channel resource utilization. Besides, RA UEs have to contend for RA channel resources in an uncoordinated manner due to no channel reservation. Therefore, making efficient use of channel resources to simultaneously support a large number of RA UEs is essential.

Recently, massive multiple input multiple output (mMIMO) %\footnote{In this paper, the terminology mMIMO refers to co-located mMIMO}
has been a key technology in 5G and future wireless communication to mitigate wireless
resource scarcity and increase channel-resource utilization efficiency \cite{6}.
As a large number of either co-located antennas (co-located mMIMO) or distributed antennas (distributed mMIMO) are employed at the base station (BS) in mMIMO, mutual channel orthogonality among RA UEs (also known as favorable propagation) can be asymptotically achieved as the number of antennas increases \cite{7,7_1}. By taking advantage of this property, RA UEs can share the same channel resource simultaneously without the need of channel reservation, while beamforming techniques can be used to spatially separate them in an effective manner.
Thus, mMIMO has been considered a prominent enabler for GFRA.

A number of research works have been undertaken to study the performance of GFRA with co-located mMIMO.
As pointed out by \cite{8}, preamble collision (i.e., multiple RA UEs choose the same preamble) is the main bottleneck in GFRA with co-located mMIMO that curbs the performance of RA UEs. In fact, since the traffic of mMTC RA UEs is usually random and sporadic, the BS has neither prior information of RA UEs' activity nor their channel state information (CSI) in each GFRA slot. Thus, each RA UE needs to send a preamble prior to data for channel estimation. However, since the number of orthogonal
preambles is limited and RA UEs choose preambles in a random and uncoordinated
manner, there could be multiple RA UEs that select the same preamble.
As a result, the estimated CSI for these collided RA UEs becomes inaccurate and data from them would be incorrectly decoded.
Various approaches have been developed to address the preamble collision issue in GFRA with co-located mMIMO. For instance, non-orthogonal preambles are considered in \cite{9} to expand preamble space without constraint on the preamble length and sporadic traffic pattern of RA UEs is exploited to detect and identify active RA UEs. In \cite{10_1}, a super-preamble consisting of multiple short preambles is adopted and features of favorable propagation and channel hardening in mMIMO are exploited to identify the super-preamble of each RA UE. In \cite{11_1}, an ensemble
independent component analysis (EICA) based pilot random access is proposed to enable joint active UEs detection and uplink data decoding. These approaches are effective in reducing the preamble collision.
In \cite{8_3}, a successive interference cancellations (SIC) based scheme is proposed for preamble collision resolution. However, this type of scheme requires that RA UEs use multiple RA slots to transmit multiple preambles and same data to make SIC possible.
Therefore, how to resolve the preamble collision when it occurs on the basis of a single RA slot is still an open issue in GFRA. In fact, since all the signals of collided RA UEs are multiplexed and assembled at the centralized BS in co-located mMIMO, the BS can only deem that the received signals come from a single RA UE, making it practically difficult to find preamble multiplicity (i.e., the number of the RA UEs that select the same preamble) and resolve the preamble collision. Note that conventional preamble decontamination solutions, such as \cite{8_1,8_2}, cannot be used to solve the considered problem in GFRA since they require that the information of the number of active UEs and all UEs' partial CSI (e.g., large scale fading coefficients) is available at the BSs as a prior knowledge.
In \cite{12_1}, an effective preamble collision resolution scheme is proposed in co-located mMIMO. Nevertheless, it only works in grant-based RA as a feedback after preamble detection from the BS to RA UEs is required.

Different from co-located mMIMO in terms of antenna topology, distributed mMIMO employs a large number of geographically distributed access points (APs) to serve UEs, and each AP is equipped with a single or a few antennas. Compared to co-located mMIMO, distributed mMIMO provides macro-diversity and has
enhanced network coverage and capacity \cite{13_1,14_1,15_1,16_1,11,18_1}. Nevertheless, the existing works that address the preamble collision issue in distributed mMIMO rely on the condition that the full or partial CSI of UEs is known at the BS \cite{8_4,8_5,20}, which is not the case in the context of GFRA. Thus, GFRA with distributed mMIMO has not been well investigated yet. On the other hand, since APs are spatially distributed in distributed mMIMO and signals to different APs undergo different levels of large-scale fading, only neighboring APs within a communication range of an UE have non-negligible channel gains \cite{16_1,11}, which implies signal spatial sparsity in distributed mMIMO \cite{10}. This feature opens up a possibility for preamble collision resolution in GFRA. Specifically, due to the sporadic traffic pattern of RA UEs, collided RA UEs could be separate in space and surrounded by different groups of APs.
If the BS is able to identify neighboring APs of a collided RA UE in GFRA and only employs the neighboring APs rather than all the APs to serve the collided RA UE, the interference from other collided RA UEs in the preamble domain could be largely mitigated and its performance is expected to be improved as a result.

Motivated by this, a novel machine learning enabled AP clustering scheme is proposed to resolve the preamble collision in GFRA by leveraging distributed mMIMO. To facilitate preamble collision resolution, collided preamble multiplicity needs to be estimated by the BS. To this end,
we first design a tailored deep neural network (DNN) to enable the preamble multiplicity estimation in GFRA, where a data-driven energy detection (ED) method is also proposed for performance comparison. With the estimated preamble multiplicity, we propose a $K$-means AP clustering algorithm to cluster the neighboring APs of collided RA UEs, and then each AP cluster is employed to decode the received data individually. Under practical wireless environments and different deployments of distributed mMIMO, we investigate and analyze the performance of the proposed DNN and show that decent estimation accuracy and reliability can be achieved.
Simulation results further confirm that the proposed machine learning enabled AP clustering schemes are able to achieve a near-optimal performance in terms of preamble collision resolution, and provide significant performance enhancement over the traditional schemes.

% suffer from different degrees of path loss caused by different access distances to distributed antenna arrays
The novelty and contribution of this paper are summarized
as follows.
\begin{itemize}
  \item We propose a novel machine learning based framework solution to mitigate the impact of preamble collision on the performance of collided RA UEs in GFRA with distributed mMIMO, which requires neither prior information of RA UEs' activity nor their CSI. To the best of our knowledge, this is the first work that aims to resolve the preamble collision in GFRA on the basis of a single RA slot by exploiting two-dimensional signal information including signal strength and locations in distributed mMIMO.
  \item To enable preamble collision resolution in GFRA, the preamble multiplicity is an indispensable parameter that needs to be estimated by the BS. To this end, we for the first time leverage deep learning based classification models to enable the preamble multiplicity estimation in distributed mMIMO, where connections between received preamble signal patterns and preamble multiplicities are exploited.
  \item With the estimated preamble multiplicity, we further propose a $K$-means AP clustering algorithm to enable the neighboring AP clustering of collided RA UEs and organize each AP cluster instead of all the APs to decode data of collided RA UEs individually. Thereby, the mutual interference among collided RA UEs in the preamble domain could be effectively mitigated, which results in appreciable performance improvement.
\end{itemize}

The remainder of this paper is organized as follows. In Section II,
the system model of GFRA with distributed mMIMO is introduced and the motivation of this work is explained theoretically by a toy example. In Section III, the proposed DNN for preamble multiplicity estimation is detailed and its estimation performance is investigated.
In Section IV, the proposed $K$-means AP clustering algorithm is presented and
its performance in terms of uplink achievable rate per collided RA UE is
evaluated. The work is concluded in Section V.
%The channel reservations are essential in the legacy LTE RA scheme since the BS is unable to accommodate multiple RA UEs simutanously over the same channel.

\normalem
\emph{Notation: }Boldface lower and upper case symbols represent
vectors and matrices, respectively. $\mathbf{I}_n$ is the $n \times n$ identity matrix. The conjugate, transpose, and complex conjugate transpose
operators are denoted by $(\cdot)^{*}$, $(\cdot)^{\mathrm{T}}$ and $(\cdot)^{\mathrm{H}}$. $\|\cdot\|$ denotes the Euclidean norm. $\mathbf{x}\sim \mathcal{CN}(0,\mathbf{\Sigma})$ indicates that $\mathbf{x}$ is a circularly symmetric
complex Gaussian (CSCG) random vector with zero-mean and
covariance matrix $\mathbf{\Sigma}$.
%problems are usually too complex to be modelled due to the dynamic wireless environments. However, dynamic patterns of the wireless environment could be effectively explored by ML with much lower complexity than using optimization technologies

\section{System Models and Motivation}

\subsection{System Model}
\begin{figure}[!h]
	%\vspace{-0.1cm}  %调整图片与上文的垂直距离
	%%\setlength{\abovecaptionskip}{-0.1cm}   % 调整图片标题与图距离
	%\setlength{\belowcaptionskip}{-0.2cm}   %调整图片标题与下文距离
	\centering
	\includegraphics[width=4in]{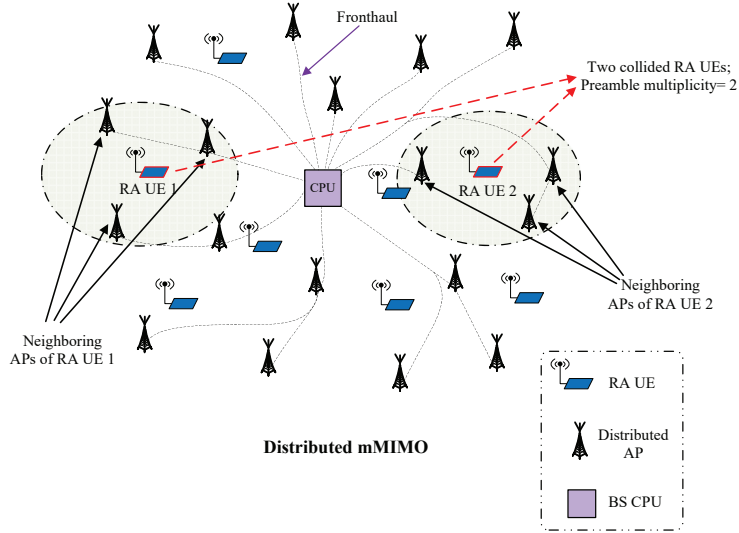}
	% where an .eps filename suffix will be assumed under latex,
	% and a .pdf suffix will be assumed for pdflatex; or what has been declared
	% via \DeclareGraphicsExtensions.
	%\captionsetup{justification=centering}
	\caption{Illustration of distributed mMIMO systems and an example with preamble multiplicity of two.} \label{fig1}
\end{figure}

We consider a distributed mMIMO system in a wide area to serve $N$ MTC UEs that are scattered
in the area (each UE is equipped with a single antenna). As illustrated in Figure \ref{fig1}, there are $M$ APs uniformly and spatially distributed. We assume that these distributed APs are connected to a BS central processing unit (CPU) via an error-free fronthaul and each AP is equipped with $S$ antennas.
%For antenna-topology comparison, a co-located mMIMO system is illustrated on the top of Figure \ref{fig1}, where the antennas are spatially co-located.
%with co-located mMIMO systems  (shown on the top of Figure \ref{fig1}),

In a GFRA channel slot\footnote{In practice, a GFRA slot consists of multiple channels over frequency and each channel can accommodate a number of RA UEs. Since each channel is independent in frequency, we only focus on a single channel scenario in this work.}, suppose that $U$ UEs, indexed as $1,2, \ldots, U$, are active to access the channel for uplink transmission in a grant-free manner, where $U$ follows the binomial distribution $\mathrm{Bino}(N, \rho)$ and $\rho$ ($\rho \ll 1$) is the sporadic activation probability of each UE.

To enable channel estimation at the APs, each RA UE directly transmits an RA preamble before data, which is randomly selected from an orthogonal preamble pool of size $L$ ($L \ll N$),
i.e., $\Omega=\{\mathbf{p}_1,\mathbf{p}_2,\ldots,\mathbf{p}_L\}$, where $\mathbf{p}_l$ denotes the $l$th orthogonal preamble vector of length $L$,
$\|\mathbf{p}_{l}\|^2=L$ and $\mathbf{p}^\mathrm{T}_{l}\mathbf{p}^{\ast}_{l'}=0$, for $l\neq l'$, $l, l' \in \{1,2,\dots,L\}$. In this paper, as in the RA of LTE, single-root Zadoff-Chu sequence is used to generate the orthogonal preambles thanks to its good auto-correlation properties \cite{3_0}.

Specifically, the received preamble signal, $\mathbf{Y}  \in \mathbb{C}^{MS\times L}$, can be given by
\begin{align}\label{eq0}
\mathbf{Y}=\sum\limits_{u=1}^{U}\sqrt{P_\mathrm{T}}\mathbf{g}_{u}\bm{\psi}^{\mathrm{T}}_u+\mathbf{N},%\underbrace{}_{\bm{\xi}_1},
\end{align}
where ${P_\mathrm{T}}$ is the transmit power of each RA UE, $\mathbf{g}_{u}=[\mathbf{g}^\mathrm{T}_{u1},\mathbf{g}^\mathrm{T}_{u2},\ldots, \mathbf{g}^\mathrm{T}_{uM}]^\mathrm{T} \in \mathbb{C}^{MS}$ is the channel response vector between RA UE $u$ and the APs and $\mathbf{g}_{um}=\sqrt{\beta_{um}}\mathbf{h}_{um} \in \mathbb{C}^{S}$ is the channel response vector between RA UE $u$ and AP $m$, where $\beta_{um}$ denotes the large-scale fading coefficient and $\mathbf{h}_{um} \sim \mathcal{CN}(0,\mathbf{I}_{S})$ stands for the small-scale fading vector, $\bm{\psi}_u \in \mathbb{C}^{L}$ is the selected preamble by RA UE $u$ from the preamble pool $\Omega$, and $\mathbf{N}$ is the noise matrix with i.i.d. elements distributed as $\mathcal{CN}(0,\sigma^2)$.

Due to the randomness of preamble selection by each RA UE, a key issue to be addressed is the preamble collision, which constrains the throughput and transmission reliability of collided RA UEs.
In the sequel, we detail the performance impairment caused by preamble collision in GFRA and present the intuition and motivation of preamble collision resolution in distributed mMIMO.

%In this paper, we aim to propose an effective approach to resolve preamble collision in distributed mMIMO and improve the achievable uplink rate of an arbitrary RA UE under preamble collision compared to that in co-located mMIMO.
%In a GFRA slot, we denote the number of active RA UEs as $K$ ($K=1,2,\ldots, N $). So there are K bursts transmitted with
%%In this paper, an independent Rayleigh block fading propagation channel model is considered. The channel response vector between the $n$th RA UE ($n=1,2,\ldots, N $) and the $m$th AP ($m=1,2,\ldots, M $) is modelled by $\mathbf{g}_{nm}=\sqrt{\beta_{nm}}\mathbf{h}_{nm} \in \mathbb{C}^{S}$, where $\beta_{nm}$ denotes the large-scale fading coefficient between $n$th RA UE and the $m$th AP, and $\mathbf{h}_{nm} \sim \mathcal{CN}(0,\mathbf{I}_{S})$ stands for the small-scale fading vector.

%assume the 1st preamble is the collided preamble that selected by the 1st RA UE.
% open avenues for preamble collision resolution and performance enhancement in GFRA.

\subsection{Performance Impairment due to Preamble Collision}\label{PIPC}
%In co-located mMIMO, since all $MS$ antennas are geographically centralized, we have $\beta_{k1}=\beta_{k2}=\cdots=\beta_{kM} \triangleq \beta_{k}$. By taking advantage of the features of channel hardening and favourable propagation, ideal preamble detection (i.e., identifying whether or not a preamble is selected by any RA UE) is asymptotically achievable as $MS \to \infty$ \cite{9}. Nevertheless, when preamble collision occurs, the BS is unable to identify the preamble multiplicity (i.e., how many RA UEs select the same preamble) and only deems that any collided preamble is only selected by a single RA UE.

Without loss of generality, we consider the RA UE with index $1$ as the RA UE of interest and explain the impact of preamble collision on its performance. In the case of preamble collision, without any prior CSI information, the least-squares (LS) based channel estimation for RA UE $1$ can be used and the estimate over all the APs is given by
\begin{align}\label{eq1}
\mathbf{\hat{g}}_1=\frac{\mathbf{Y}\bm{\psi}^{*}_1}{\sqrt{P_\mathrm{T}}L}=\mathbf{g}_{1}+\sum\limits_{u' \in \Phi_{\bm{\psi}_1}}\mathbf{g}_{u'}+\frac{1}{\sqrt{\rho_\mathrm{T}L}}{\mathbf{{n}}},%\underbrace{}_{\bm{\xi}_1},
\end{align}
where $\Phi_{\bm{\psi}_1}$ is the set of indices of RA UEs that select $\bm{\psi}_1$ other than RA UE 1, the cardinality of $\Phi_{\bm{\psi}_1}$ is denoted by $|\Phi_{\bm{\psi}_1}|\geq1 $, $\rho_{\mathrm{T}} = P_{\mathrm{T}}/\sigma^2$ is defined as the uplink transmit signal-to-noise ratio (SNR) corresponding to each RA UE, and $\mathbf{{n}} \sim \mathcal{CN}(0,\mathbf{I}_{MS})$. From (\ref{eq1}), we see that the estimated channel under preamble collision is distorted by the channels of other RA UEs that select the same preamble.

Following preamble, each RA UE transmits its data. The received data symbol vector $\mathbf{r} \in \mathbb{C}^{MS}$ over all the APs is given by
\begin{align}\label{eq2}
\mathbf{r}=\sum_{u=1}^{U}\sqrt{P_\mathrm{T}}\mathbf{g}_us_u+\bar{\mathbf{n}},
\end{align}
where $\bar{\mathbf{n}}$ the background noise vector distributed as $\mathcal{CN}(0,\sigma^2\mathbf{I}_{MS})$ and $s_{u} $ is a data symbol transmitted by RA UE $u$ and $\mathbb{E}[|s_{u}|^2]=1$.

Then, all the APs send their channel estimate and received data signals to the BS CPU. Since the error-free fronthaul links between all the APs and the BS CPU are assumed as in \cite{16_1,11,18_1,8_4,8_5,20, 10}, the information sent by APs can be perfectly gathered at the BS CPU for signal processing.
With (\ref{eq1}) and (\ref{eq2}), the estimated data symbol of RA UE 1 after conjugate beamforming at the BS CPU is thus given by
\begin{align}\label{eq2_1}
\hat{s}_{1}=&\frac{\mathbf{\hat{g}}_1^{\mathrm{H}}\mathbf{r}}{MS\sqrt{P_\mathrm{T}}}
= \frac{\mathbf{\hat{g}}_1^{\mathrm{H}}\mathbf{g}_1s_1}{MS}+\frac{\sum_{u=2}^{U}\mathbf{\hat{g}}_1^{\mathrm{H}}\mathbf{g}_us_u}{MS}
+\frac{\mathbf{\hat{g}}_1^{\mathrm{H}}\bar{\mathbf{n}}}{MS\sqrt{P_\mathrm{T}}}.
\end{align}
As $M \to \infty$ (here we fix $S$), it becomes
\begin{align}\label{eq3}
\hat{s}_{1\infty}=&\lim_{M \to \infty}\left(\frac{\mathbf{\hat{g}}_1^{\mathrm{H}}\mathbf{g}_1s_1}{MS}+\frac{\sum_{u=2}^{U}\mathbf{\hat{g}}_1^{\mathrm{H}}\mathbf{g}_us_u}{MS}
+\frac{\mathbf{\hat{g}}_1^{\mathrm{H}}\bar{\mathbf{n}}}{MS\sqrt{P_\mathrm{T}}}\right) \nonumber \\
\overset{(a)}{=} & \frac{\mathbf{{g}}_1^{\mathrm{H}}\mathbf{g}_1}{MS}s_{1}+\sum\limits_{u' \in \Phi_{\bm{\psi}_1}}\frac{\mathbf{{g}}_{u'}^{\mathrm{H}}\mathbf{g}_{u'}}{MS}s_{u'} \nonumber\\
\overset{(b)}{=} & \lim_{M \to \infty}\sum\limits_{m=1}^{M}\frac{\beta_{1m}s_1}{M}+\sum\limits_{u' \in \Phi_{\bm{\psi}_1}}\lim_{M \to \infty}\sum\limits_{m=1}^{M}\frac{\beta_{u'm}s_{u'}}{M} \nonumber\\
=&\beta_{1}s_1 + \sum\limits_{u' \in \Phi_{\bm{\psi}_1}}\beta_{u'}s_{u'},
\end{align}
where $\overset{(a)}{=}$ and $\overset{(b)}{=}$ are obtained based on Chebyshev's Theorem. Specifically, $\overset{(a)}{=}$ is obtained by the fact that $\frac{\mathbf{{g}}_u^{\mathrm{H}}\mathbf{g}_{u'}}{SM} \xrightarrow[M \to \infty]{P} 0$ when $u \ne u' $ and $\frac{\mathbf{\hat{g}}_1^{\mathrm{H}}\bar{\mathbf{n}}}{SM} \xrightarrow[M \to \infty]{P} 0$. $\overset{(b)}{=}$ is obtained by the fact that, $\frac{\mathbf{{g}}_u^{\mathrm{H}}\mathbf{g}_u}{S}=\sum\limits_{m=1}^{M}\frac{\beta_{um}\|\mathbf{h}_{um}\|^2}{S} $. Since $\frac{\|\mathbf{h}_{um}\|^2}{S}$ follows the gamma distribution with shape $S$ and scale $\frac{1}{S}$, it has mean of $1$ and variance of $\frac{1}{S}$. Thus, we have
$\frac{\mathbf{{g}}_u^{\mathrm{H}}\mathbf{g}_u}{MS} \xrightarrow[M \to \infty]{P} %\stackrel{M \to \infty}{\longrightarrow}
 \frac{\sum\limits_{m=1}^{M}\beta_{um}}{M}$.
In addition, $\beta_{u}\triangleq \lim\limits_{M \to \infty}\sum\limits_{m=1}^{M}\frac{\beta_{um}}{M}$, $u=1,2,\ldots,U$.

As a result, the asymptotic signal-to-interference-and-noise ratio (SINR) of RA UE 1 is expressed by
\begin{align}\label{eq4}
\mathrm{SINR}_{1\infty}
&=\frac{ \beta^2_{1}}{\sum\limits_{u' \in \Phi_{\bm{\psi}_1}}\beta^2_{u'}}.
\end{align}
%Note that since the BS has no any prior CSI of RA UEs, $\mathrm{SINR}_{1\infty}$ is a genie-aided SINR expression that only used to facilitate understand the impact of preamble collision.
As we can see, even the number of antennas increases without bound, it does not change the fact that the interference from the collided RA UEs that select the same preamble as RA UE 1 cannot be vanished and could have a significant impact on the performance of RA UE 1.

In co-located mMIMO, since all $M$ APs are geographically centralized, we have $\beta_{u1}=\beta_{u2}=\cdots=\beta_{uM} \triangleq \beta_{u}$ and thus the observation in (\ref{eq4}) still holds. Unfortunately, in co-located mMIMO, since all the signals are multiplexed and assembled at the centralized BS, it is difficult to find preamble multiplicity under preamble collision and the performance impairment of collided RA UEs exists no matter where collided RA UEs are spatially located. However, distributed mMIMO opens up chances for mitigating the impairment thanks to the signal spatial sparsity in distributed mMIMO and random geographic distributions of RA UEs.
%the feature that signals of RA UEs are geographically distributed over APs.

%By taking advantage of the features of channel hardening and favourable propagation, ideal preamble detection (i.e., identifying whether or not a preamble is selected by any RA UE) is asymptotically achievable as $MS \to \infty$ \cite{9}. Nevertheless, when preamble collision occurs, the BS is unable to identify the preamble multiplicity (i.e., how many RA UEs select the same preamble) and only deems that any collided preamble is only selected by a single RA UE.

%hat{x}_{1}=&\mathbf{\hat{g}}_1^{\mathrm{H}}\mathbf{r} \nonumber \\
%=&\sqrt{P_\mathrm{T}}\mathbf{\hat{g}}_1^{\mathrm{H}}\mathbf{\hat{g}}_1x_1 \nonumber \\
%&+\sqrt{P_\mathrm{T}}\mathbf{\hat{g}}_1^{\mathrm{H}}(\mathbf{g}_1-\mathbf{\hat{g}}_1)x_1 +\sum_{k=2}^{K}\sqrt{P_\mathrm{T}}\mathbf{\hat{g}}_1^{\mathrm{H}}\mathbf{g}_kx_k+\mathbf{\hat{g}}_1^{\mathrm{H}}\bar{\mathbf{n}},
%where $\sqrt{P_\mathrm{T}}\mathbf{\hat{g}}_1^{\mathrm{H}}\mathbf{\hat{g}}_1x_1$ is treated as the useful signal, and the other three terms on the right side is the self-interference, multiuser interference, and noise, respectively. As a result, the SINR of RA UE 1 can be written as

%
%\begin{align}\label{eq4}
%\mathrm{SINR}^C_{1}
%&=\frac{\rho_{\mathrm{T}}\|\mathbf{\hat{g}}_1\|^4}{\rho_{\mathrm{T}}|\mathbf{\hat{g}}_1^{\mathrm{H}}(\mathbf{g}_1-\mathbf{\hat{g}}_1)|^2 + \rho_{\mathrm{T}}\sum_{k=2}^{K}|\mathbf{\hat{g}}_1^{\mathrm{H}}\mathbf{g}_k|^2 + \|\mathbf{\hat{g}}_1\|^2}.
%\end{align}

\subsection{Preamble Collision Resolution in Distributed mMIMO based GFRA}
 \label{PCR}
In distributed mMIMO, considering the distance disparity between an RA UE and different APs, it is demonstrated that only neighboring APs within a communication range of an RA UE have non-negligible channel gains. Since all the collided RA UEs are uniformly and independently distributed in the area, they can be separate in space and surrounded by different groups of APs.
%\footnote{Since the size of closed region of a UE is comparably smaller than that of a cell, the probability that collided RA UEs that select the same preamble are located geographically in the close vicinity and sharing the same adjacent antennas is significantly lower than the probability that they located separately and have different adjacent antennas.}.
If the BS CPU can distinguish the neighboring APs of a collided RA UE in GFRA, it can organize the neighboring APs to serve the collided RA UE, which is expected to improve the performance of collided RA UEs significantly.

Herein, we use a toy example to explain the potential performance gain achieved by such a strategy.
In particular, we assume that RA UE 1 is far away from the RA UEs in $\Phi_{\bm{\psi}_1}$ so that the strength of received signals from the other collided RA UEs is negligible at the neighboring APs of RA UE $1$ (for example in Figure \ref{fig1}, RA UE $1$ and RA UE $2$ are the collided RA UEs that select the same preamble but their locations are far away from each other). For simplicity, let $\mathcal{M}_1=[1,2,\ldots, M_1]$ denote the set of indices of neighboring APs of RA UE 1 and ${M}_1=|\mathcal{M}_1|=\omega_1 M$, where $\omega_1$ ($0<\omega_1\ll 1$) is a scaling factor that represents the ratio of the sizes of a communication range of RA UE $1$ to the considered area.

By only employing the ${M}_1$ APs to decode data, similar to (\ref{eq1}), the channel estimate of RA UE $1$ over the ${M}_1$ APs, $\mathbf{\hat{g}}_{1,\mathcal{M}_1} \in \mathbb{C}^{{M}_1S}$, can be written by
\begin{align}\label{eq5}
\mathbf{\hat{g}}_{1,\mathcal{M}_1} =& \mathbf{g}_{1,\mathcal{M}_1}+\sum\limits_{u' \in \Phi_{\bm{\psi}_1}}\mathbf{g}_{u',\mathcal{M}_1}+\frac{1}{\sqrt{\rho_\mathrm{T}L}}{\mathbf{{n}}_{\mathcal{M}_1}}, \end{align}
where $\mathbf{g}_{u,\mathcal{M}_1}=[\mathbf{g}^\mathrm{T}_{u1},\mathbf{g}^\mathrm{T}_{u2},\ldots,\mathbf{g}^\mathrm{T}_{uM_1}]^\mathrm{T}$ and $\mathbf{{n}_{\mathcal{M}_1}} \sim \mathcal{CN}(0,\mathbf{I}_{M_1S})$.

Similar to (\ref{eq2}), the received data symbol vector over the ${M}_1$ APs, $ \mathbf{r}_{\mathcal{M}_1} \in \mathbb{C}^{{M}_1S}$, is written by
\begin{align}\label{eq6}
\mathbf{r}_{\mathcal{M}_1}=&\sum_{u=1}^{U}\sqrt{P_\mathrm{T}}\mathbf{g}_{u,\mathcal{M}_1}s_u+\bar{\mathbf{n}}_{\mathcal{M}_1},
\end{align}
where $ \bar{\mathbf{n}}_{\mathcal{M}_1} \sim \mathcal{CN}(0,\sigma^2\mathbf{I}_{M_1S})$.

In the considered example, due to significant large-scale fading between RA UE $u'$ in $\Phi_{\bm{\psi}_1}$ and AP $m$ in $\mathcal{M}_1$, $\mathbf{g}_{u',\mathcal{M}_1} \approx \mathbf{0}, u'\in \Phi_{\bm{\psi}_1}$. Thus, we have following approximations:
\begin{align*}
\mathbf{\hat{g}}_{1,\mathcal{M}_1}
 \approx \mathbf{g}_{1,\mathcal{M}_1}+\frac{1}{\sqrt{\rho_\mathrm{T}L}}{\mathbf{{n}}_{\mathcal{M}_1}},%\underbrace{}_{\bm{\xi}_1},
\end{align*}
and
\begin{align*}
\mathbf{r}_{\mathcal{M}_1}
 \approx \sqrt{P_\mathrm{T}}\mathbf{g}_{1,\mathcal{M}_1}s_1+\sum_{\substack{u\notin \Phi_{\bm{\psi}_1}\\ u\neq 1 }}\sqrt{P_\mathrm{T}}\mathbf{g}_{u,\mathcal{M}_1}s_u+\bar{\mathbf{n}}_{\mathcal{M}_1}.
\end{align*}
Then, the estimated data symbol of RA UE $1$ after conjugate beamforming as $M \to \infty$ (as $M_1=\omega_1 M$, $M \to \infty$ leads to $M_1 \to \infty$) becomes
\begin{align}\label{eq7}
\hat{s}_{1\infty}=&\lim_{M_1 \to \infty}\frac{\mathbf{\hat{g}}_{1,\mathcal{M}_1}^{\mathrm{H}}\mathbf{r}_{\mathcal{M}_1}}{M_1S\sqrt{P_\mathrm{T}}} \nonumber \\
\overset{(b)}{\approx} & \lim_{M_1 \to \infty}\sum\limits_{m=1}^{M_1}\frac{\beta_{1m}s_1}{M_1} =\beta_{1,\mathcal{M}_1}s_1.
\end{align}
Similarly, $\overset{(b)}{\approx}$ is obtained based on Chebyshev's Theorem as $M_1 \to \infty$ and $\beta_{1,\mathcal{M}_1}\triangleq\lim\limits_{M_1 \to \infty}\sum\limits_{m=1}^{M_1}\frac{\beta_{1m}}{M_1}$.

From (\ref{eq7}), we can see that the received signal of RA UE 1 approximately becomes interference-free from preamble collision under the given scenario as $M \to \infty$, which indicates the possibility and practicality of preamble collision resolution (mitigating the interference due to preamble collision) in GFRA with distributed mMIMO.

Nevertheless, to achieve the potential preamble collision resolution and improve the performance of collided RA UEs in GFRA by distributed mMIMO, there are two issues remained to be addressed as follows:
\begin{itemize}
  \item How to detect preamble collision and find preamble multiplicity?
  \item How to differentiate the neighboring APs of collided RA UEs for performance enhancement?
\end{itemize}

To address the above issues, it is expected to fully exploit the information obtained from the received preamble signals at APs. To this end, we propose a machine learning based framework solution in this paper.

Specifically, to mitigate the performance impairment of collided RA UEs in GFRA with distributed mMIMO, we first design a simple DNN to enable the preamble multiplicity estimation, where a data-driven ED method is also proposed for performance comparison. With the estimated preamble multiplicity, we then employ the $K$-means clustering algorithm to separate the neighboring APs of collided RA UEs and use each associated AP cluster to serve individual collided RA UE.

\section{DNN based Preamble Multiplicity Estimation}
For the estimation of preamble multiplicity of an arbitrary preamble, e.g., $\mathbf{p}_l$, $l =1,2,\ldots, L$, we need to find out the mapping relationship between the preamble multiplicity and the received preamble signal associated with $\mathbf{p}_l$ at APs, i.e.,
\begin{align}\label{eq8}
 F \colon & \mathbb{C}^{MS}  \parbox{.5cm}{\rightarrowfill} \mathbb{N}_0 \nonumber\\
 & \mathbf{g}_{\mathcal{B}_{l}}\mapsto {B}_l,~l =1,2,\ldots, L
\end{align}
where $F$ denotes the mapping function, $\mathbb{N}_0=\mathbb{N} \cup \{0\}$, $\mathcal{B}_{l}$ is the set of indices of RA UEs that select $\mathbf{p}_l$ among $U$ RA UEs and ${B}_l=|\mathcal{B}_{l}| \in \mathbb{N}_0$
denotes the preamble multiplicity (e.g., ${B}_l=0$ indicates that $\mathbf{p}_l$ is not selected by any RA UE),
and $\mathbf{g}_{\mathcal{B}_{l}} \in \mathbb{C}^{MS}$ represents the received preamble signal associated with $\mathbf{p}_l$ at APs, which has a similar expression as in (\ref{eq1}) and it is given by
\begin{align}\label{eq9}
 \mathbf{g}_{\mathcal{B}_{l}}&=\frac{\mathbf{Y}\mathbf{p}^{*}_1}{\sqrt{P_\mathrm{T}}L}
 =\sum\limits_{u \in \mathcal{B}_{l}}\mathbf{g}_{u}+\frac{1}{\sqrt{\rho_\mathrm{T}L}}{\mathbf{{n}}},
\end{align}
where $\mathbf{g}_{\mathcal{B}_{l}}=[\mathbf{g}^\mathrm{T}_{{\mathcal{B}_{l}}1},\mathbf{g}^\mathrm{T}_{{\mathcal{B}_{l}}2},\ldots,\mathbf{g}^\mathrm{T}_{{\mathcal{B}_{l}}M}]^\mathrm{T}$
and $\mathbf{g}_{{\mathcal{B}_{l}}m}$ is the received preamble signal vector associated with $\mathbf{p}_l$ at AP $m$.

In the considered problem, obtaining $F$ by traditional programming algorithms is not a trivial task since deriving a general mathematical detection model to recognize subtle patterns associated with different preamble multiplicities and summarize
the random patterns of RA UEs' geographic locations and wireless environments in GFRA
is too complex and may be infeasible. %learn meaningful features and

\subsection{Proposed DNN Structure}
\begin{figure}[!h]
	%\vspace{-0.1cm}  %调整图片与上文的垂直距离
	%%\setlength{\abovecaptionskip}{-0.1cm}   % 调整图片标题与图距离
	%\setlength{\belowcaptionskip}{-0.2cm}   %调整图片标题与下文距离
	\centering
	\includegraphics[width=4in]{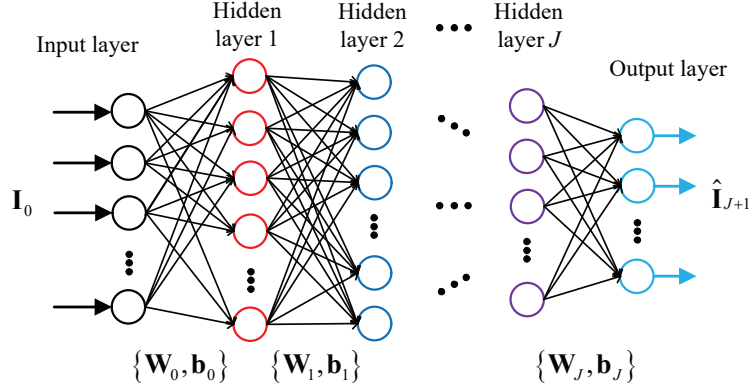}
	% where an .eps filename suffix will be assumed under latex,
	% and a .pdf suffix will be assumed for pdflatex; or what has been declared
	% via \DeclareGraphicsExtensions.
	%\captionsetup{justification=centering}
	\caption{A simplified illustration of the proposed DNN diagram for preamble multiplicity estimation, where the circle nodes of different colors represent the neurons of different layers.} \label{fig2}
\end{figure}

To solve the problem in an effective manner, we design a feed-forward DNN (multi-layer perception) \cite{13} in this section thanks to its powerful approximation and prediction ability. As aforementioned in Section \ref{PCR}, only neighboring APs within a communication range of an RA UE have non-negligible channel gains in distributed mMIMO. Thus, the neighboring APs of an RA UE usually capture more significant signal energy than the other APs.
On the other hand, the separation of RA UEs in space makes the signals of different RA UEs concentrate in different geographic clusters.
By capitalizing these properties, one could envision that there exists connection between the preamble multiplicities and the distribution patterns of received preamble signal energy over $M$ distributed APs. Therefore, the proposed DNN is used to explore the features so that the desired function $F$ can be approximately modelled.

As illustrated in Figure \ref{fig2}, the proposed fully connected DNN consists of $J+2$ layers, including one input layer (layer 0), $J$ hidden layers (layers 1 to $J$), and one output layer (layer $J+1$). Let $N_j$ denote the number of neurons at layer $j$, $j=0,1,\ldots, J+1$.

In the proposed DNN, layer 0 contains $N_0=M$ neurons, which forwards the instantaneous information of $\mathbf{g}_{\mathcal{B}_{l}}$ to the following layers. Let $\mathbf{E}_{\mathcal{B}_{l}}=[E_{{\mathcal{B}_{l}}1},E_{{\mathcal{B}_{l}}2},\ldots,E_{{\mathcal{B}_{l}}M}]^{\mathrm{T}} \in \mathbb{R}^{M}$ denote the received preamble signal energy vector associated with $\mathbf{p}_l$, where
\begin{align}\label{eq10}
 E_{{\mathcal{B}_{l}}m}=\frac{\|\mathbf{g}_{{\mathcal{B}_{l}}m}\|^2}{S}, m=1,2,\ldots, M,
\end{align}
denotes the received preamble signal energy associated with $\mathbf{p}_l$ at AP $m$.
With $\mathbf{E}_{\mathcal{B}_{l}}$, the input vector of layer 0, denoted by $\mathbf{I}_0 =[I_{01}, \ldots, I_{0m}, \ldots, I_{0M}]^\mathrm{T} \in \mathbb{R}^{M}$, is given by
\begin{align}\label{eq11}
\mathbf{I}_0=\mathrm{sort}_\mathrm{D}(\mathbf{E}_{\mathcal{B}_{l}}),
\end{align}
where $\mathrm{sort}_\mathrm{D}(\cdot)$ is a function that sorts the elements in descending order.

\begin{figure}[!h]
	%\vspace{-0.1cm}  %调整图片与上文的垂直距离
	%%\setlength{\abovecaptionskip}{-0.1cm}   % 调整图片标题与图距离
	%\setlength{\belowcaptionskip}{-0.2cm}   %调整图片标题与下文距离
	\centering
	\includegraphics[width=4in]{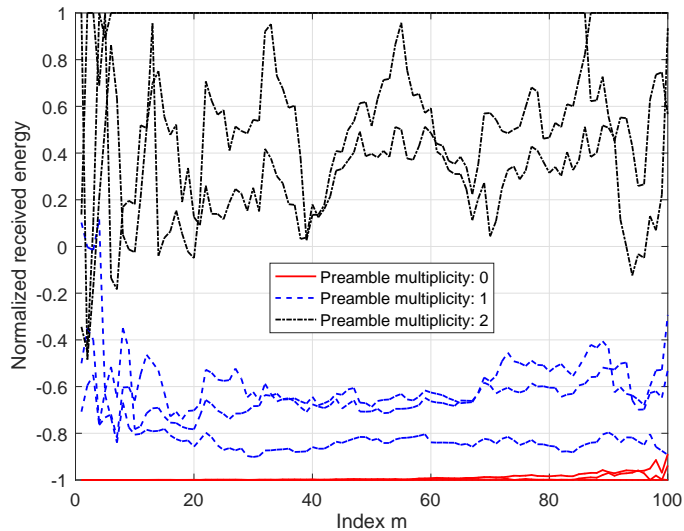}
	% where an .eps filename suffix will be assumed under latex,
	% and a .pdf suffix will be assumed for pdflatex; or what has been declared
	% via \DeclareGraphicsExtensions.
	%\captionsetup{justification=centering}
	\caption{An example of pattern differences of normalized $\mathbf{I}_0$ corresponding to different preamble multiplicities with $M=100$, where the x-label is the index of the elements in $\mathbf{I}_0$.} \label{fig21}
\end{figure}

In Figure \ref{fig21}, we plot an example to illustrate the pattern features of normalized $\mathbf{I}_0$ corresponding to three different preamble multiplicities with $M=100$, where we randomly generate three sample sets for each individual preamble multiplicity based on the system setup in Section \ref{setup} and perform normalization among them. As shown in the example, different preamble multiplicities lead to different received energy patterns, which are exploited by the proposed DNN to predict preamble multiplicity. Note that we only use nine sample sets in Figure \ref{fig21} to illustrate the received energy pattern differences of different preamble multiplicities. In practice, the received energy patterns become more complicated as more sample sets are involved for normalization.%Therefore, are hard to

%Based on the universal approximation theorem stated in \cite{12}, DNN is able to approximate any non-linear function, up to arbitrary accuracy.

Building on $\mathbf{I}_0$, the hidden layers of the feed-forward DNN are constructed through the following $J$ iterative processing steps:
\begin{align}\label{eq12}
\mathbf{I}_j=f(\mathbf{W}_{j-1}\mathbf{I}_{j-1}+\mathbf{b}_{j-1}), j=1,2,\ldots, J,
\end{align}
where $\mathbf{I}_j \in \mathbb{R}^{N_j}$ is the output of layer $j$, $f(\cdot)$ represents a non-linear activation function,
and $\mathbf{W}_{j-1} \in \mathbb{R}^{N_j \times N_{j-1}}$  and $\mathbf{b}_{j-1} \in \mathbb{R}^{N_j}$ respectively stand for the weighting matrix and bias vector at layer $j-1$, which are used to encode the output of layer $j-1$. In this paper, a sigmoid function defined as $\sigma(x)=\frac{1}{1+e^{-x}}$ is used as the activation function $f(\cdot)$.

As usually done in pattern recognition problems, the predicted output of the proposed DNN in layer $J+1$ is expressed by
\begin{align}\label{eq13}
\mathbf{\hat{I}}_{J+1}=\mathrm{softmax}(\mathbf{W}_{J}\mathbf{I}_{J}+\mathbf{b}_{J}),
\end{align}
where $\mathbf{\hat{I}}_{J+1}
=[\hat{I}_{J+1,1},\hat{I}_{J+1,2},\ldots, \hat{I}_{J+1,N_{J+1}}]^{\mathrm{T}} \in \mathbb{R}^{N_{J+1}}$ and $\mathrm{softmax}(\cdot)$ is the softmax function \cite{13}. In the proposed DNN, $N_{J+1}$ is set to ${T}_\mathrm{max} + 1$, where $T_\mathrm{max}$ is the maximum number of colliding RA UEs that we are
interested in detecting. Thus, the output $\mathbf{\hat{I}}_{J+1}$ represents a predicted probability distribution over the $T_\mathrm{max}+1$ different possible preamble multiplicities and the value of $\hat{I}_{J+1,i}$ indicates the predicted probability that the preamble multiplicity equals $i-1$, $i=1,2,\ldots,T_\mathrm{max}+1$.

In the light of this, the estimated preamble multiplicity of $\mathbf{p}_l$ according to the output of proposed DNN is given by
\begin{align}\label{eq14}
\hat{B}_l=\operatorname*{arg\,max}(\mathbf{\hat{I}}_{J+1})-1,
\end{align}
where $\operatorname*{arg\,max}(\mathbf{\hat{I}}_{J+1})$ returns the index of the largest entry of $\mathbf{\hat{I}}_{J+1}$.

\subsection{Training Phase}
For an accurate approximation of the desired function $F$, the proposed DNN needs to learn and adjust the parameter sets of $\bm{\theta}_j=\{\mathbf{W}_{j}, \mathbf{b}_{j}\}, j=0, 1,\ldots,J$, in the training phase.

Specifically, based on the system configurations in GFRA, we randomly generate $Q$ training sample sets.
 For sample set $q$ ($ q=1,2, \ldots, Q$), it consists of a pair of the received preamble signal energy $\mathbf{E}_{\mathcal{B}^q}$ and the corresponding preamble multiplicity $B^q$, which is associated with an arbitrary preamble (here we omit the subscript $l$ for notation simplicity). With known training sample set $q$, we can obtain a mapping pair of the proposed DNN, denoted by $(\mathbf{I}^q_{0}, \mathbf{I}^q_{J+1})$, where the input
$\mathbf{I}^q_{0}$ is obtained from $\mathbf{E}_{\mathcal{B}^q}$ by (\ref{eq11}), and the target output $\mathbf{I}^q_{J+1}$ can be expressed by
 \begin{align}\label{eq15}
 \mathbf{I}^q_{J+1}=\mathbf{e}_{B^q+1},
 \end{align}
where $\mathbf{e}_i \in \mathbb{R}^{T_\mathrm{max}+1}$ denotes the standard basis vector that has a single nonzero entry with value 1 at entry $i$.

By using the $Q$ mapping pairs, the proposed DNN is trained by back-propagation (BP) algorithm to adjust and optimize the parameter sets of all layers, i.e., $\bm{\theta}=[\bm{\theta}_0,\bm{\theta}_1,\ldots, \bm{\theta}_J]$, so that the cross-entropy loss function between the target outputs $\{\mathbf{I}^q_{J+1}\}_{q=1}^{Q}$ and the predicted outputs
$\{\mathbf{\hat{I}}^q_{J+1}\}_{q=1}^{Q}$, which is given in (\ref{eq16}), could be minimized:
 \begin{align}\label{eq16}
\mathcal{L}(\bm{\theta})=-\sum\limits_{q=1}^{Q}\sum\limits_{i=1}^{T_\mathrm{max}+1}{{I}}^q_{J+1,i}\ln\left({\hat{I}}^q_{J+1,i}(\mathbf{I}^q_{0}, \bm{\theta})\right).
 \end{align}

To achieve the minimization of $\mathcal{L}(\bm{\theta})$, a number of out-of-the-box gradient methods including the gradient descent and the conjugate gradient can be used. In this paper, we employ the scaled conjugate gradient method \cite{15} to iteratively update parameter sets $\bm{\theta}$ \cite{14}. %out-of-the-box training algorithms

\subsection{Performance Analysis of Proposed DNN}\label{DNNperf}
\subsubsection{Simulation Setup}\label{setup}
In this subsection, we evaluate the performance of the proposed DNN for preamble multiplicity estimation under practical wireless environments, where the large-scale fading coefficient, depending on the RA UE's location and the propagation environment, is modelled as \cite{19},
 \begin{align}\label{eq17}
\beta_{um}=\frac{X_{um}}{1+\mathrm{PL}(d_0)\left(\frac{d_{um}}{d_0}\right)^v},
 \end{align}
where $X_{um}$ stands for the the shadow
fading that is a log-normal random variable with standard deviation $\sigma_{\mathrm{SF}}$ (dB),  $\mathrm{PL}(d_0)$ is the path loss at a reference distance $d_0$ (m), $d_{um}$ is the distance between UE $u$ and AP $m$, and $v$ is the path loss exponent. The additive
thermal noise is assumed to have a power spectral density of
$-174$ dBm/Hz, while the front-end receiver at the AP is assumed to have a noise figure of $9$ dB according to \cite{20}.
Thus, the noise power $\sigma^2$ is $-112$ dBm with a narrow bandwidth of $B_\mathrm{w}=200$ KHz.

We consider a square area of 1 km$^2$ and the distributed APs are deployed on a square grid\footnote{Note that the real AP deployments could be irregular due to network planning and other topological and demographic factors. However, the APs cannot be randomly distributed in practice because it may lead to APs being placed very close to each other, which generally does not make sense. Thus, APs are typically distributed more regularly than randomly distributed. For this reason, the widely accepted square-grid-based AP deployment model is considered in this paper. Moreover, this paper takes into account the shadow fading. The randomness caused by shadow fading coefficients can be seen as displacing the APs and varying the distances between the APs and the RA UEs \cite{11}. Therefore, the considered model is a sensible model for AP deployments.}.
Three different deployments are considered: 1) $M=10 \times 10 $ APs with $S=2$ antenna; 2) $M=10 \times 10 $ APs with $S=1$ antenna; and 3) $M=7\times7$ APs with $S=2$ antennas.
The rest of system parameters are summarized in Table \ref{table1}.
\begin{table}[htp]
\small
	\renewcommand{\arraystretch}{1.4}
	\caption{System parameters}\label{table1} \centering
	\begin{tabular}{>{\centering}m{2cm}|>{\centering}m{2cm}|>{\centering}m{3cm}|}
		%\hline\hlinem
		%\multicolumn{2}{|c|}  {\multirow{1}*{}}
		%\multicolumn{3}{|c|}{\multirow{1}*{}}
		\hline
		%\cline
		
		\multicolumn{1}{|c|}{Number of UEs $N$}  & \multicolumn{2}{c|}{$2000$} \tabularnewline
		\hline
		
		\multicolumn{1}{|c|}{Activation Probability $\rho$}  & \multicolumn{2}{c|}{$0.01$} \tabularnewline
		\hline
		
		\multicolumn{1}{|c|}{Number of Preambles $L$}  & \multicolumn{2}{c|}{$20$} \tabularnewline
		\hline
		
		\multicolumn{1}{|c|}{Transmit Power $P_\mathrm{T}$}  & \multicolumn{2}{c|}{$17$ dBm} \tabularnewline
		
		\hline
		\multicolumn{1}{|c|}{Shadow Fading $\sigma_{\mathrm{SF}}$}  & \multicolumn{2}{c|}{$0$ or $8$ dB} \tabularnewline
        \hline
		\multicolumn{1}{|c|}{Reference Distance $d_0$}  & \multicolumn{2}{c|}{$1$ m} \tabularnewline
		\hline
		\multicolumn{1}{|c|}{Path Loss $\mathrm{PL}(d_0)$}  & \multicolumn{2}{c|}{$30$ dB} \tabularnewline
		\hline
		\multicolumn{1}{|c|}{Path Loss Exponent $v$}  & \multicolumn{2}{c|}{$3.8$} \tabularnewline
		\hline

	\end{tabular}
\end{table}

For the proposed DNN, we simulate $Q=10^5$ realizations to generate sample sets. To improve generalization and avoid overfitting, the sample sets are randomly divided into training sample sets (80\% of total instances), validation sample sets (10\% of total instances), and test sample sets (10\% of total instances). Moreover, the minimum performance gradient is set to be $10^{-6}$. The maximum number of epochs to train is set to be $1000$. And the maximum number of validation checks is set to be $8$.
For each realization, the active RA UEs are randomly distributed in the considered area and their number $U$ is generated following the binomial distribution $\mathrm{Bino}(N, \rho)$. Since each preamble is selected by RA UEs uniformly at random in GFRA, the preamble multiplicity $B$ associated with an arbitrary preamble follows the binomial distribution $\mathrm{Bino}(N, \rho/L)$ as mentioned earlier. With the given system parameters, i.e., $N=2000$, $\rho=0.01$, and $L=20$, over 99\% realizations are generated in the way that a preamble is selected by $4$ RA UEs at most. As a result, we set $T_\mathrm{max}=4$ as the maximum preamble multiplicity that we are interested in estimating. In the following, we discuss the performance of the proposed DNN in terms of classification accuracy and reliability in different deployments.

\subsubsection{Performance Analysis for Different Deployments} %$1$: $M=10 \times 10 $ with $S=2$
\label{PADD}
We first consider the performance of deployment $1$ with $M=10 \times 10 $ and $S=2$. In this deployment, the proposed DNN consists of 4 hidden layers, whose numbers of neurons are $128$, $128$, $64$, and $32$, respectively.

\begin{table}[h!]
\small
    \centering
    \renewcommand\arraystretch{1.35}
     \renewcommand{\multirowsetup}{\centering}

\caption{Confusion matrix for the proposed DNN in the deployment of $M=100 $ and $S=2$.}\label{table2}
\setlength{\tabcolsep}{1.5mm}{
    \begin{tabular}{@{} ccc| ccccc}
      \multicolumn{1}{c}{} & \multicolumn{1}{c}{} &\multicolumn{1}{c}{} &\multicolumn{5}{c}{\bf{Predicted} $\hat{B}$} \\

\multicolumn{1}{c}{} & \multicolumn{1}{c}{} &
\multicolumn{1}{c|}{}     & 0 & 1 & 2 & 3 & 4       \\
        \cline{2-8}
     %\multirow{4}{Actual}% {*}{\rotatebox[origin=tr]{90}{Actual}} %{\rothead {Actual Class}}
       \multirow{10}{1cm}{{\bf{Target} $B$}} & \multirow{5}{1cm}{$\sigma_{\mathrm{SF}}=0$ dB}
       & 0     & \bf{1}       &  0       & 0      & 0     & 0   \\
     &   & 1     & 0.002  &  \bf{0.997}   & 0.001  & 0     & 0   \\
     &   & 2     & 0      &  0.012   & \bf{0.979}  & 0.009 & 0    \\
      &  & 3     & 0      &  0       & 0.067  & \bf{0.918} & 0.015    \\
      &  & 4     & 0      &  0       & 0.002  & 0.120 & \bf{0.878 }   \\
           \cline{2-8}
           \cline{2-8}
    &  \multirow{5}{1cm}{$\sigma_{\mathrm{SF}}=8$ dB}   & 0     & \bf{1}       &  0       & 0      & 0     & 0   \\
      &  & 1     & 0.002  &  \bf{0.991}   & 0.007  & 0     & 0   \\
      &  & 2     & 0      &  0.046   & \bf{0.923}  & 0.031 & 0    \\
      &  & 3     & 0      &  0       & 0.151  & \bf{0.838} & 0.011    \\
       & & 4     & 0      &  0       & 0.002    & 0.211 & \bf{0.787 }   \\
           \cline{2-8}
    \end{tabular}}
\end{table}

To understand the classification accuracy of the proposed DNN for preamble multiplicity, confusion matrices for different $\sigma_{\mathrm{SF}}$ in Table \ref{table2} are included. In the scenario with no shadow fading, i.e., $\sigma_{\mathrm{SF}}=0$, it is seen that the proposed DNN model is able to predict (estimate) the preamble multiplicity in GFRA with high accuracy over a wide range of multiplicities. In terms of the performance of preamble detection, i.e., determining whether a preamble is selected or not by any RA UE, the proposed DNN model provides almost error-free performance, with negligible false alarm and missed detection errors ($0.2$\% missed detection probability occurs merely when preamble is selected by only one RA UE). When the preamble collision occurs, about $98$\% and $88$\% estimation accuracy can be achieved when the preamble multiplicity equals $2$ and $4$, respectively. The main reason that the estimation accuracy declines as the preamble multiplicity increases is due to the fact that, a larger preamble multiplicity means that more collided RA UEs select the same preamble. As a consequence, there are comparably more chances that some of the collided RA UEs are co-located in vicinity, which could make the proposed DNN mistakenly treat these close-located RA UEs as a single RA UE and results in an incorrect estimated preamble multiplicity that is smaller than the actual one. Therefore, the estimation performance is degraded. Nevertheless, it is noticed that, almost all the incorrect estimated multiplicities are only offset by $1$ compared to the actual ones. For example, when the preamble multiplicity equals $2$ and $3$, the proposed DNN only gets the multiplicity wrong by $\pm 1$ (mostly by $-1$). When the preamble multiplicity equals $4$, the proposed DNN guarantees an estimation result that is either correct or incorrect by $\pm 1$ with a high probability of $99.8$\% (only gets the multiplicity incorrect by $-2$ with a as little as $0.2$\% probability). These observations demonstrate the accuracy as well as the reliability achieved by the proposed DNN for preamble multiplicity estimation.

In addition, we also consider a practical channel scenario with shadow fading $\sigma_{\mathrm{SF}}=8$. Under such a condition, it is not surprising that, due to the impact of shadow fading variations on channel gains, the classification accuracy of the proposed DNN degrades compared to the case without shadow fading. As we can see, although shadow fading has little impact on the preamble detection performance of the proposed DNN, it incurs certain accuracy degradations for estimating collided preamble multiplicities. For instance, the estimation accuracy for a preamble multiplicity of $4$ is decreased from $87.8$\% to $78.7$\% and more errors are introduced by incorrect estimation to multiplicity $3$, which indicates that the channel randomness induced by shadow fading inherently increases confusion between adjacent multiplicity classes.
Nevertheless, an estimation accuracy of $78.7$\% for preamble multiplicity $4$ is still considered decent, under such an amount of collided RA UEs coexists at the same time. Besides, similar to what we observed in the case with no shadow fading, almost all the incorrect estimated multiplicities differ from the true ones by $\pm 1$ when $\sigma_{\mathrm{SF}}=8$, which reveals that although the accuracy performance of the proposed DNN is affected by the shadow fading, its estimation reliability remains uninfluenced.

%the proposed DNN is designed and trained built on the randomness of spatial distributions of RA UEs in distributed mMIMO. The more collided RA UEs are separate, the pattern disparities of received signal distributions over APs between different preamble multiplicities are more obvious. As a result,

%\subsubsection{Performance in Deployment $2$: $M=10 \times 10 $ with $S=1$}

\begin{table}[h!]
\small
    \centering
    \renewcommand\arraystretch{1.35}
     \renewcommand{\multirowsetup}{\centering}
    \settowidth\rotheadsize{\theadfont Actual Class}
\caption{Confusion matrix for the proposed DNN in the deployment of $M=100 $ and $S=1$.}\label{table3}
\setlength{\tabcolsep}{1.5mm}{
    \begin{tabular}{@{} ccc| ccccc}
       \multicolumn{1}{c}{} &\multicolumn{1}{c}{} &\multicolumn{1}{c}{} &\multicolumn{5}{c}{\bf{Predicted} $\hat{B}$} \\

\multicolumn{1}{c}{} &\multicolumn{1}{c}{} &
\multicolumn{1}{c|}{}     & 0 & 1 & 2 & 3 & 4       \\
        \cline{2-8}
     %\multirow{4}{Actual}% {*}{\rotatebox[origin=tr]{90}{Actual}} %{\rothead {Actual Class}}
       \multirow{10}{1cm}{\bf{Target}\\ $B$} & \multirow{5}{1cm}{$\sigma_{\mathrm{SF}}=0$ dB}
       & 0     & \bf{1}       &  0       & 0      & 0     & 0   \\
      &  & 1     & 0.004  &  \bf{0.992}   & 0.004  & 0     & 0   \\
      &  & 2     & 0      &  0.038   & \bf{0.943}  & 0.019 & 0    \\
      &  & 3     & 0      &  0       & 0.116  & \bf{0.864} & 0.020    \\
      &  & 4     & 0      &  0       & 0.002  & 0.218 & \bf{0.780 }   \\
           \cline{2-8}
           \cline{2-8}

     %\multirow{4}{Actual}% {*}{\rotatebox[origin=tr]{90}{Actual}} %{\rothead {Actual Class}}

      &  \multirow{5}{1cm}{$\sigma_{\mathrm{SF}}=8$ dB} & 0     & \bf{1}       &  0       & 0      & 0     & 0   \\
       & & 1     & 0.006  &  \bf{0.981}   & 0.013  & 0     & 0   \\
       & & 2     & 0      &  0.051   & \bf{0.919}  & 0.030 & 0    \\
      &  & 3     & 0      &  0       & 0.174  & \bf{0.786} & 0.040    \\
      &  & 4     & 0      &  0       & 0.004  & 0.269 & \bf{0.727 }   \\
           \cline{2-8}
    \end{tabular}}
\end{table}

We also consider other two deployments, i.e., deployment $2$ with $M=10 \times 10 $ and $S=1$ and deployment $3$ with $M=7 \times 7 $ and $S=2$. Their confusion matrices are illustrated in Table \ref{table3} and Table \ref{table4}, respectively. In deployment $2$, the proposed DNN consists of $4$ hidden layers, whose numbers of neurons are the same as those in deployment $1$. In deployment $3$, the proposed DNN consists of $4$ hidden layers, whose numbers of neurons are $64$, $128$, $64$, and $32$, respectively. As observed in Table \ref{table2}, similar observations and conclusions can be drawn from Tables \ref{table3} and \ref{table4}. In terms of estimation accuracy of the proposed DNN, we can see that it is slightly degraded in deployment $2$ compared to that in deployment $1$, which is mainly due to a loss of channel diversity in deployment $2$ with $S=1$. Nevertheless, a $72.7$\% estimation accuracy for preamble multiplicity $4$ is still achievable with $\sigma_{\mathrm{SF}}=8$ in deployment $2$. Moreover, with the roughly same amount of antennas in deployments $2$ and $3$, a reasonably close multiplicity estimation performance is observed without considering shadow fading. However, results show that the performance in deployment $3$ seems more sensitive to the channel randomness resulted from shadow fading. In particular, under $\sigma_{\mathrm{SF}}=8$, its estimation accuracy for preamble multiplicities $3$ and $4$ is significantly degraded compared to that under $\sigma_{\mathrm{SF}}=0$. This could be explained by the fact that the antenna distribution in deployment $3$ is more sparse, which makes that the shadow fading comparably has more significant impact on the channel fluctuations. As a result, the classification confusion between preamble multiplicities $3$ and $4$ gets more pronounced.

%\subsubsection{Performance in Deployment 3: $M=7\times7$ with $S=2$}

\begin{table}[h!]
\small
    \centering
    \renewcommand\arraystretch{1.5}
     \renewcommand{\multirowsetup}{\centering}
\caption{Confusion matrix for the proposed DNN in the deployment of $M=49 $ and $S=2$.}\label{table4}
\setlength{\tabcolsep}{1.5mm}{
    \begin{tabular}{@{} ccc| ccccc}
       \multicolumn{1}{c}{} &\multicolumn{1}{c}{} &\multicolumn{1}{c}{} &\multicolumn{5}{c}{\bf{Predicted} $\hat{B}$} \\

\multicolumn{1}{c}{} &\multicolumn{1}{c}{} &
\multicolumn{1}{c|}{}     & 0 & 1 & 2 & 3 & 4       \\
        \cline{2-8}
     %\multirow{4}{Actual}% {*}{\rotatebox[origin=tr]{90}{Actual}} %{\rothead {Actual Class}}
       \multirow{10}{1cm}{\bf{Target}\\ $B$}& \multirow{5}{1cm}{$\sigma_{\mathrm{SF}}=0$ dB}
       & 0     & \bf{1}       &  0       & 0      & 0     & 0   \\
     &   & 1     & 0.003  &  \bf{0.994}   & 0.003  & 0     & 0   \\
      &  & 2     & 0      &  0.034   & \bf{0.946}  & 0.020 & 0    \\
     &   & 3     & 0      &  0       & 0.138  & \bf{0.830} & 0.032    \\
     &   & 4     & 0      &  0       & 0.002  & 0.233 & \bf{0.765 }   \\
           \cline{2-8}

        \cline{2-8}
     %\multirow{4}{Actual}% {*}{\rotatebox[origin=tr]{90}{Actual}} %{\rothead {Actual Class}}

    &  \multirow{5}{1cm}{$\sigma_{\mathrm{SF}}=8$ dB}   & 0     & \bf{1}       &  0       & 0      & 0     & 0   \\
      &  & 1     & 0.003  &  \bf{0.971}   & 0.026  & 0     & 0   \\
      &  & 2     & 0      &  0.061   & \bf{0.924}  & 0.015 & 0    \\
      &  & 3     & 0      &  0       & 0.313  & \bf{0.663} & 0.024    \\
      &  & 4     & 0      &  0       & 0.018  & 0.396 & \bf{0.586 }   \\
           \cline{2-8}
    \end{tabular}}
\end{table}

\subsection{Comparison to Threshold-Based ED Method}\label{T-ED}
To further validate the effectiveness of the proposed DNN based method for preamble multiplicity estimation, we consider a threshold-based ED method for performance comparison in this subsection. As indicated in Figure \ref{fig21}, different preamble multiplicities, to some extent, lead to different levels of received energy after normalization among training sample sets. For this reason, a simple threshold-based ED (T-ED) method for preamble multiplicity estimation can be developed by the following steps:
\begin{enumerate}
	\item
	Perform energy normalization among $Q$ training sample sets between $-1$ and $1$ and store the normalization setting.
    For sample set $q$ ($ q=1,2, \ldots, Q$), its normalized energy set is denoted by $\tilde{\mathbf{I}}^q_0$.
    \item %Calculate average normalized energy per AP for each multiplicity: \\
	Denote $\mathcal{V}_{B}$ as the set of indices of sample sets associated with preamble multiplicity $B$ ($B=0,1,...$). Calculate the average normalized energy per AP for preamble multiplicity $B$ as $\zeta_B=\frac{\sum_{q \in \mathcal{V}_{B}}\|\tilde{\mathbf{I}}^q_0\|_1}{M|\mathcal{V}_{B}|}$, where $\|\cdot\|_1$ is the Taxicab norm.
    \item Obtain threshold $\mathrm{Th}_B$ for preamble multiplicity estimation, where $\mathrm{Th}_0=-1$ and
    $\mathrm{Th}_B=\frac{\zeta_B+\zeta_{B+1}}{2}$ for $B\neq0$.
    \item For a new received energy set $\mathbf{I}_0$, apply the normalization setting to it and obtain its average normalized energy per AP as $\frac{\|\tilde{\mathbf{I}}_0\|_1}{M}$. If $\mathrm{Th}_B< \frac{\|\tilde{\mathbf{I}}_0\|_1}{M}\leq \mathrm{Th}_{B+1}$, its preamble multiplicity is estimated as $B$.
\end{enumerate}
%\begin{enumerate}
%	\item
%	Perform energy normalization among $Q$ training sample sets between $-1$ and $1$ and store the normalization setting.
%    For sample set $q$ ($ q=1,2, \ldots, Q$), its normalized energy set is denoted by $\tilde{\mathbf{I}}^q_0$.
%    \item %Calculate average normalized energy per AP for each multiplicity: \\
%	Denote $\mathcal{V}_{B}$ as the set of indices of sample sets associated with preamble multiplicity $B$ ($B=0,1,...$). Calculate the average normalized energy per AP for preamble multiplicity $B$ as $\zeta_B=\frac{\sum_{q \in \mathcal{V}_{B}}\|\tilde{\mathbf{I}}^q_0\|_1}{M|\mathcal{V}_{B}|}$, where $\|\cdot\|_1$ is the Taxicab norm.
%    \item Obtain threshold $\mathrm{Th}_B$ for preamble multiplicity estimation, where $\mathrm{Th}_0=-1$ and
%    $\mathrm{Th}_B=\frac{\zeta_B+\zeta_{B+1}}{2}$ for $B\neq0$.
%    \item For a new received energy set $\mathbf{I}_0$, apply the normalization setting to it and obtain its average normalized energy per AP as $\frac{\|\tilde{\mathbf{I}}_0\|_1}{M}$. If $\mathrm{Th}_B< \frac{\|\tilde{\mathbf{I}}_0\|_1}{M}\leq \mathrm{Th}_{B+1}$, its preamble multiplicity is estimated as $B$.
%\end{enumerate}
\begin{table}[h!]
\small
    \centering
    \renewcommand\arraystretch{1.35}
     \renewcommand{\multirowsetup}{\centering}

\caption{Confusion matrix of the T-ED method in deployment $1$ with $\sigma_{\mathrm{SF}}=8$.}\label{table5}
\setlength{\tabcolsep}{1.5mm}{
    \begin{tabular}{@{} cc| ccccc}
      \multicolumn{1}{c}{} &\multicolumn{1}{c}{} &\multicolumn{5}{c}{\bf{Predicted} $\hat{B}$} \\

\multicolumn{1}{c}{} & \multicolumn{1}{c|}{}     & 0 & 1 & 2 & 3 & 4      \\
        \cline{2-7}
     %\multirow{4}{Actual}% {*}{\rotatebox[origin=tr]{90}{Actual}} %{\rothead {Actual Class}}
       \multirow{5}{1cm}{\bf{Target}\\ $B$}
        &0     & \bf{1}       &  0       & 0      & 0     & 0    \\
       & 1     & 0.107  &  \bf{0.892}   & 0.001  & 0     & 0   \\
       & 2     & 0      &  0.131   & \bf{0.844}  & 0.025 & 0    \\
       & 3     & 0      &  0       & 0.155  & \bf{0.764} & 0.081    \\
       & 4     & 0      &  0       & 0  & 0.281 & \bf{0.719 } \\

    \end{tabular}}
\end{table}
Like the proposed DNN, the developed T-ED method is a data-driven based method. For performance comparison, its confusion matrix in deployment $1$ with $\sigma_{\mathrm{SF}}=8$ is presented in Table \ref{table5}. Comparing the results to the ones in Table \ref{table2}, we see that the estimation performance of the T-ED method is not as good as that of the proposed DNN. This is because that the T-ED method is unable to exploit the received energy patterns over APs of different preamble multiplicities as the proposed DNN does. Since similar conclusions can be drawn as in deployment $1$, we omit the confusion matrices of the T-ED method in deployments $2$ and $3$. In Section V, performance comparison between the DNN based scheme and the T-ED based scheme in terms of uplink achievable rate per collided RA UE will be illustrated in Figure \ref{fig6}.

\textit{Discussion}: For the proposed DNN, training with the setup of $M=100$ requires approximately $400$ iterations, which in total take around $10$ minutes. The training time becomes shorter with a smaller $M$. We note that the training cost is incurred offline and the cost burden of the proposed DNN is on the BS CPU, whose capabilities are rapidly improving with artificial intelligence (AI) processors and with little computational and energy restrictions. Thus, the incurred cost and computational complexity may not be a limiting factor. Besides, the training frequency depends on the environment stability, e.g., stability of the placement and the number of APs. Since the placement and the number of APs over a covered area should remain unchanged over a long period of time in practice, the training should be performed infrequently.

In the next section, the estimated preamble multiplicity information will be used to cluster neighboring APs of collided RA UEs for their performance enhancement.

\section{Proposed AP Clustering Algorithm}
The estimated preamble multiplicity $\hat{B}$ (associated with an arbitrary preamble) based on the %$\mathbf{p}_l$
proposed DNN indicates the status of associated preamble in GFRA, i.e., whether or not it is selected by any RA UE, and if selected then how many RA UEs select it. When $\hat{B} \geq 2$, the BS CPU assumes that preamble collision occurs. Under such conditions, as revealed in Section \ref{PCR},
it is expected that the BS CPU only allocates the neighboring APs of a collided RA UE (rather than all the APs) to decode its data so that the mutual interference among collided RA UEs in the preamble domain can be mitigated. In the light of this, we propose a $K$-means AP clustering algorithm in this section.

\subsection{$K$-Means AP Clustering Algorithm}
In this paper, we denote $M_\mathrm{c}$ as the average number of neighboring APs to decode for each collided RA UE in the case of preamble collision. %\footnote{Note that, since this work aims to shed light on the employment of DNN techniques for preamble collision resolution in distributed mMIMO, the optimization of $M_\mathrm{c}$ is out of scope and we assume that $M_\mathrm{c}$ is a pre-configured system parameter in this paper.}
Ideally, the neighboring APs of a collided RA UE can be the $M_\mathrm{c}$ APs with its strongest channel gains \cite{17}. In practice, this scenario is desirable, but unattainable in GFRA since the BS CPU has no prior CSI of RA UEs.
As a compromised solution, the $K$-means AP clustering algorithm is proposed to cluster neighboring APs for collided RA UEs.

On one hand, the neighboring APs in the vicinity of an RA UE usually capture more significant signal energy than other APs. As the collided RA UEs are randomly distributed in space, it can be reasonably envisaged that the APs with $M_\mathrm{c}\hat{B}$ strongest received preamble energy are most likely composed by the neighboring APs of $\hat{B}$ collided RA UEs.
On the other hand, the $K$-means clustering algorithm is one of the most popular clustering algorithms, which aims to partition observations into $K$ clusters where each observation belongs to exactly one cluster with the nearest mean cluster centroid \cite{18}. For these reasons, it motivates us to propose the $K$-means AP clustering algorithm that iteratively partitions the APs corresponding to largest $M_\mathrm{c}\hat{B}$ entries of $\mathbf{E}_{\mathcal{B}}$ into $\hat{B}$ clusters based on their coordinates. Note that the deployment of distributed APs along with their coordinates are pre-determined and known at the BS CPU.

Herein, we denote $\mathcal{A}_{\hat{B}}$ as the set of indices of APs corresponding to the largest $M_\mathrm{c}\hat{B}$ entries of $\mathbf{E}_{\mathcal{B}}$ and $|\mathcal{A}_{\hat{B}}|=M_\mathrm{c}\hat{B}$.
Then, we have $\mathcal{C}_{\hat{B}}=\{\mathbf{c}_m \mid m\in \mathcal{A}_{\hat{B}} \}$ as the coordinate set of the APs in $\mathcal{A}_{\hat{B}}$, where $\mathbf{c}_m=[x_m, y_m]^\mathrm{T} $ denotes the coordinate of AP $m$, $m\in \mathcal{A}_{\hat{B}}$, in a 2-dimensional Euclidean space.

With ${\mathcal{C}}_{\hat{B}}$, the proposed $K$-means AP clustering algorithm is described in Algorithm \ref{A1}.

\begin{algorithm}
  \caption{Proposed $K$-means AP clustering algorithm}\label{A1}
    \hspace*{\algorithmicindent} {\textbf{Input:} $M_\mathrm{c}$, $\hat{B}$, and $\mathcal{C}_{\hat{B}}$};\\
  \hspace*{\algorithmicindent} {\textbf{Output:} A set of $\hat{B}$ clusters, i.e., $\mathcal{Z}_k=\{m \mid z_m=k, m\in \mathcal{A}_{\hat{B}}\}$, $k=1,2,\ldots,\hat{B}$};\\
    \hspace*{\algorithmicindent} {\textbf{Initialization:} Randomly select $\hat{B}$ coordinates from $\mathcal{C}_{\hat{B}}$ as the initial cluster centroids $\bm{\mu}_1,\bm{\mu}_2,\ldots, \bm{\mu}_{\hat{B}}$};
    \begin{algorithmic}[1]
  %\Procedure{MyProcedure}{}

    \Repeat
      \State{AP assignment, i.e, assign each AP in ${\mathcal{C}}_{\hat{B}}$ to its closest cluster centroid with label: $z_m={\operatorname*{arg\,min}}_k\|\mathbf{c}_m- \bm{\mu}_k\|^2$};
      \State{Update the cluster centroids, i.e., compute the mean coordinates of APs assigned in each cluster to obtain new cluster centroid: $\bm{\mu}_k=\frac{\sum\limits_{m=1}^{M_\mathrm{c}\hat{B}}\mathbb{1}(z_m=k)\mathbf{c}_m}{\sum\limits_{m=1}^{M_\mathrm{c}\hat{B}}\mathbb{1}(z_m=k)}$ };
    %\state  asas

    \Until{\{Cluster centroids are stabilized\}}
%%    \Procedure{MyProcedure}{$x,y$}
%
 % \EndProcedure
    \end{algorithmic}
    \end{algorithm}

With the AP clusters $\{\mathcal{Z}_k\}_{k=1}^{\hat{B}}$, the BS CPU deems that there exists one collided RA UE in the vicinity of each AP cluster, and organizes each cluster to decode the received data individually.

\subsection{Exemplary Outputs of Proposed Algorithm}
\begin{figure}[t]
\centering
\subfigure[With correct preamble multiplicity estimation.]
{
\centering
\includegraphics[width=4in]{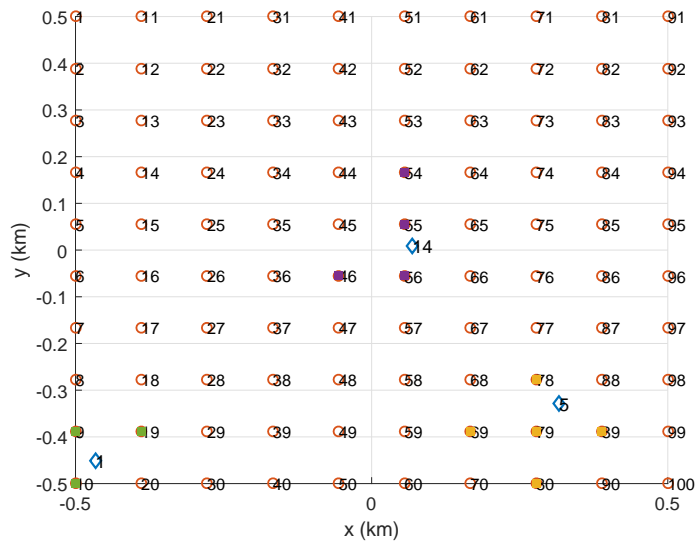}
%\caption{fig1}
\label{fig3}
}
\quad
\subfigure[With incorrect preamble multiplicity estimation.]
{
 \centering
 \includegraphics[width=4in]{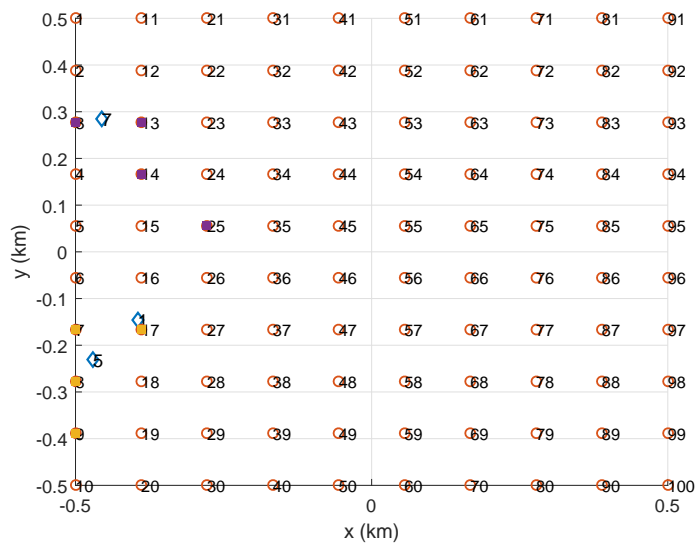}     %
 %\caption{fig1}
 \label{fig4}
}%
 \centering
 \caption{Exemplary outputs of Algorithm \ref{A1} with $M_\mathrm{c}=4$, in deployment $2$ with $\sigma_{\mathrm{SF}}=8$ dB.}
 %\label{fig10}
\end{figure}
With a predetermined $M_\mathrm{c}$, the outcome of the proposed AP clustering algorithm relies on the estimated preamble multiplicity of proposed DNN.
As observed and discussed in Section \ref{DNNperf}, the proposed DNN is able to provide a decent estimation accuracy for each preamble multiplicity. In most error multiplicity estimation, the proposed DNN only gets the multiplicity wrong by $-1$. For instance, for preamble multiplicity $3$ in deployment $2$ with $\sigma_{\mathrm{SF}}=8$ dB, an estimation accuracy of $78.6$\% is achieved and an estimation error of $17.4$\% is caused by mistakenly classifying it as multiplicity $2$.
Based on these facts, we present two kinds of representative outcomes of the proposed AP clustering algorithm with $M_\mathrm{c}=4$ in Figure \ref{fig3} and Figure \ref{fig4}, respectively. Specifically, under deployment $2$ with $\sigma_{\mathrm{SF}}=8$ dB, the outcome in Figure \ref{fig3} represents a typical clustering output for correctly estimated preamble multiplicity $3$, while the outcome in Figure \ref{fig4} represents a typical clustering output for incorrectly estimated preamble multiplicity $3$.
In both subfigures, red empty circles represent the locations of $M=100$ deployed APs in a 2-dimensional Euclidean space,
blue diamonds represent the locations of collided RA UEs (associated with a certain preamble) that are randomly generated, and the APs filled with same color form an AP cluster outputted by the proposed AP clustering algorithm. Note that there exists other RA UEs in space, however, since their signals are orthogonal to those of the displayed collided RA UEs in the preamble domain, we omit them herein.

As shown in Figure \ref{fig3}, RA UEs $1$, $5$, and $14$ select the same preamble over the same GFRA channel. With the correct preamble multiplicity estimation, the proposed AP clustering algorithm divides the $12$ APs with the strongest received preamble signal energy into $3$ clusters. Since the three collided RA UEs are geographically separate, different collided RA UEs are surrounded by different clusters of APs, each of which represents a group of APs capturing the strongest signals from a specific collided RA UE. Using each individual AP cluster rather than all the APs to decode the data of the collided RA UE in its vicinity is beneficial since the mutual interference due to preamble collision is significantly discarded.
In Figure \ref{fig4}, RA UEs $1$, $5$, and $7$ select the same preamble. Since RA UEs $1$ and $5$ are located in vicinity to each other, they are treated as a single collided RA UE by the proposed DNN, which thus mistakenly estimates the preamble multiplicity as $2$. With the incorrect preamble multiplicity estimation, the proposed AP clustering algorithm divides the $8$ APs with the strongest received preamble signal energy into $2$ clusters. In this example, RA UEs $1$ and $5$ share the same AP cluster. Although it is unable to mitigate the mutual interference between them, the interference from RA UE $7$ is mitigated by only using their neighboring clustered APs rather than all the APs. On the other hand, for RA UE $7$, the interference from the other two RA UEs becomes small at the clustered APs surrounding it and using them only to decode its data is thus beneficial.

In fact, the selection of $M_\mathrm{c}$ is of particular
importance in the proposed $K$-means AP clustering. On one hand, since the signals of an RA UE are usually concentrated at its few neighboring APs, selecting too large $M_\mathrm{c}$ brings little benefit of increasing the amount of desired signals, but introduces comparably large interference from all of the other RA UEs in GFRA. On the other hand, selecting too small $M_\mathrm{c}$ could have little benefit of further mitigating interference, but provides comparably small amount of desired signals due to limited array gain. Therefore, a proper value of $M_\mathrm{c}$ in the proposed $K$-means AP clustering needs to be selected. %balance the tradeoff between the received amount of interference and the amount of desired signals.
Unfortunately, it is difficult to determine $M_\mathrm{c}$ in a closed-form expression in the considered scenario. In this paper, we shed light on its proper value through simulation by evaluating the performance in terms of the 95\% likely achievable rate per collided RA UE.

\section{Simulation Results}
In this section, performance evaluation on the uplink achievable rate per collided RA UE is conducted by simulation to validate the effectiveness of the proposed AP clustering schemes in preamble collision resolution in GFRA. The simulation setup and parameters are the same as given in Section \ref{setup}.
%\subsection{Performance Evaluation on Achievable Rate per Collided RA UE}

%To show the advantage of the proposed DNN based AP clustering scheme for preamble collision resolution,
%We consider the uplink achievable rate of an arbitrary RA UE under preamble collision as a performance metric.
Like in Section \ref{PCR}, we consider RA UE $1$ as the collided RA UE of interest under preamble collision ($|\Phi_{\bm{\psi}_1}|\geq 1$) and $\mathcal{M}_1$ as the set of indices of APs employed for decoding data of RA UE $1$.
Based on (\ref{eq5}) and (\ref{eq6}), the estimated data symbol of RA UE $1$ is given by
\begin{align}\label{eq18}
\hat{s}_{1}=&\frac{\mathbf{\hat{g}}_{1,\mathcal{M}_1}^{\mathrm{H}}\mathbf{r}_{\mathcal{M}_1}}{M_1S\sqrt{P_\mathrm{T}}} \nonumber \\
=& \frac{\mathbf{\hat{g}}_{1,\mathcal{M}_1}^{\mathrm{H}}\mathbf{g}_{1,\mathcal{M}_1}s_1}{M_1S}
+\frac{\sum\limits_{u' \in \Phi_{\bm{\psi}_1}}\mathbf{\hat{g}}_{1,\mathcal{M}_1}^{\mathrm{H}}\mathbf{g}_{u',\mathcal{M}_1}s_u'}{M_1S}
+\frac{\sum\limits_{\substack{u\notin \Phi_{\bm{\psi}_1}\\ u\neq 1 }}\mathbf{\hat{g}}_{1,\mathcal{M}_1}^{\mathrm{H}}\mathbf{g}_{u,\mathcal{M}_1}s_u}{M_1S}
+\frac{\mathbf{\hat{g}}_{1,\mathcal{M}_1}^{\mathrm{H}}\bar{\mathbf{n}}_{\mathcal{M}_1}}{M_1S\sqrt{P_\mathrm{T}}}.
\end{align}
From (\ref{eq18}), the uplink achievable rate of RA UE $1$ under preamble collision is given by
\begin{align}\label{eq19}
\mathcal{R}_{1}=B_\mathrm{w}\log\left(1+\mathrm{SINR}_{1}\right),
\end{align}
where %$\eta$ is the transmission signalling overhead for channel estimation. Here we set $\eta=0.9$ \cite{20}.
$\mathrm{SINR}_{1}$ is the uplink SINR of RA UE $1$, which is given by \cite{17}
\begin{small}
\begin{align}\label{eq20}
\mathrm{SINR}_{1}=
\frac{\rho_{\mathrm{T}}|\mathbf{\hat{g}}_{1,\mathcal{M}_1}^{\mathrm{H}}\mathbf{g}_{1,\mathcal{M}_1}|^2}{\sum\limits_{u' \in \Phi_{\bm{\psi}_1}}\rho_{\mathrm{T}}|\mathbf{\hat{g}}_{1,\mathcal{M}_1}^{\mathrm{H}}\mathbf{g}_{u',\mathcal{M}_1}|^2\!+\!\sum\limits_{\substack{u\notin \Phi_{\bm{\psi}_1}\\ u\neq 1 }}\rho_{\mathrm{T}}|\mathbf{\hat{g}}_{1,\mathcal{M}_1}^{\mathrm{H}}\mathbf{g}_{u,\mathcal{M}_1}|^2\!+\!|\mathbf{\hat{g}}_{1,\mathcal{M}_1}^{\mathrm{H}}\bar{\mathbf{n}}_{\mathcal{M}_1}|^2}.
\end{align}
\end{small}
%Note that since the BS has no any prior CSI of RA UEs, $\mathrm{SINR}_{1}$ is a genie-aided SINR expression that used here to facilitate understand the impact of preamble collision.
%The selection of value of $M_\mathrm{c}$ is out of scope and we will show that the change of $M_1$ in a certain range has a trivial impact on the performance.
To show the performance superiority of the proposed DNN based $K$-means AP clustering scheme in terms of $\mathcal{R}_{1}$ in GFRA, simulation results are presented in the sequel. Throughout the simulations, the three deployments in Section \ref{setup} are considered only with $\sigma_{\mathrm{SF}}=8$, and the following five schemes are compared:
\begin{itemize}
  \item DNN based $K$-means AP clustering scheme: in this proposed scheme, the AP cluster closest to RA UE $1$ is employed as set $\mathcal{M}_1$.
\item T-ED based $K$-means AP clustering scheme: different from the DNN based scheme, the T-ED method developed in Section \ref{T-ED} is used for preamble multiplicity estimation.
  \item All-AP scheme: in this scheme, without preamble multiplicity estimation, all $M$ APs are employed as set $\mathcal{M}_1$.
  \item $M_\mathrm{c}$-strongest-AP scheme: in this scheme, without preamble multiplicity estimation, the $M_\mathrm{c}$ APs with the strongest received preamble signal energy are simply employed as set $\mathcal{M}_1$.

  \item Genie-aided scheme: in this genie-aided scheme, we assume that the set of APs that have the $M_\mathrm{c}$ largest channel gains of RA UE $1$ is perfectly known at the BS CPU and these APs are employed as set $\mathcal{M}_1$. The performance of this scheme can be seen as an upper bound of the proposed scheme, which is desirable but unattainable in practical GFRA.
\end{itemize}

\begin{figure*}[!t]
\centering

\includegraphics[width=2.3in]{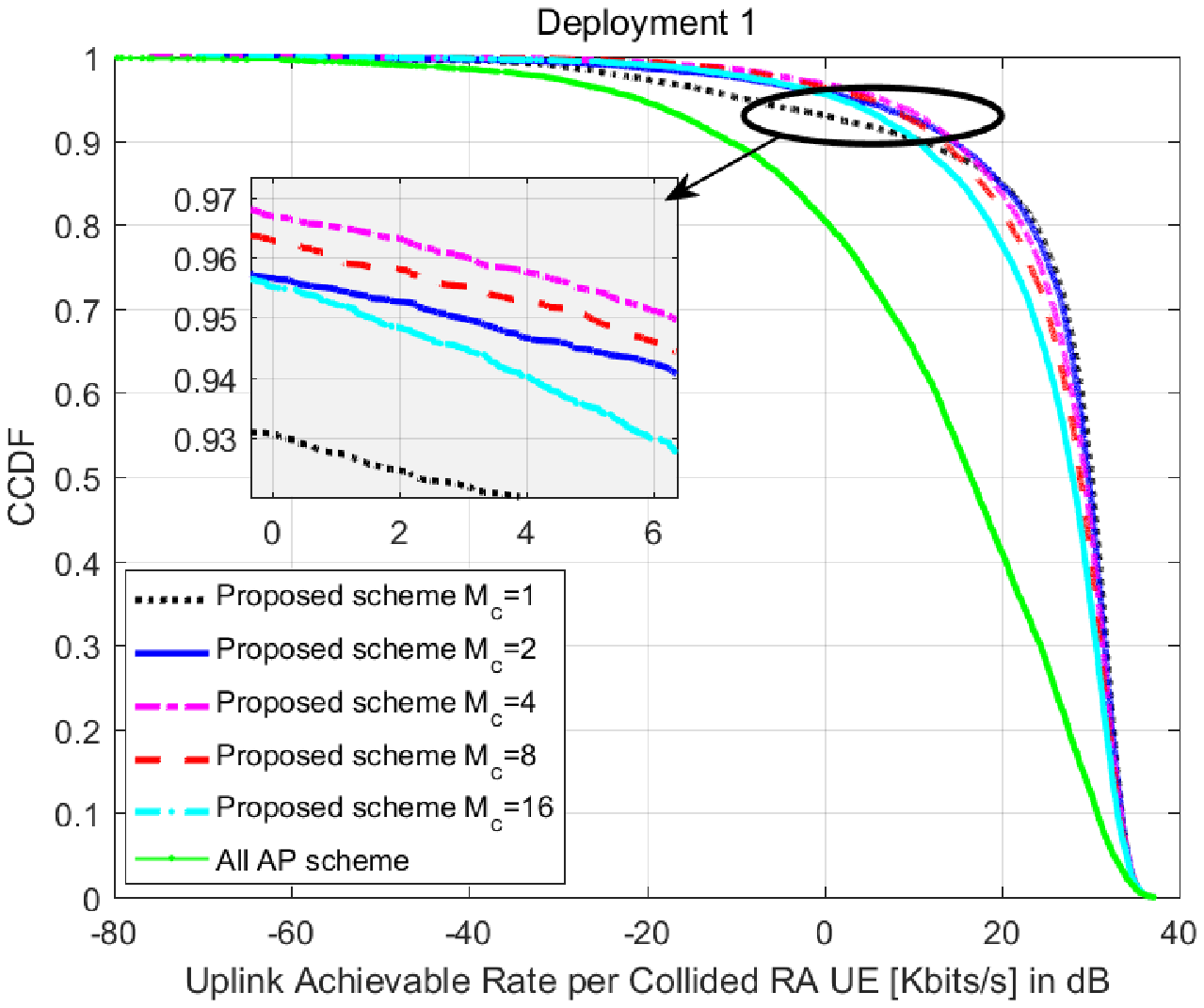}
\hspace{-4.1ex}
\includegraphics[width=2.3in]{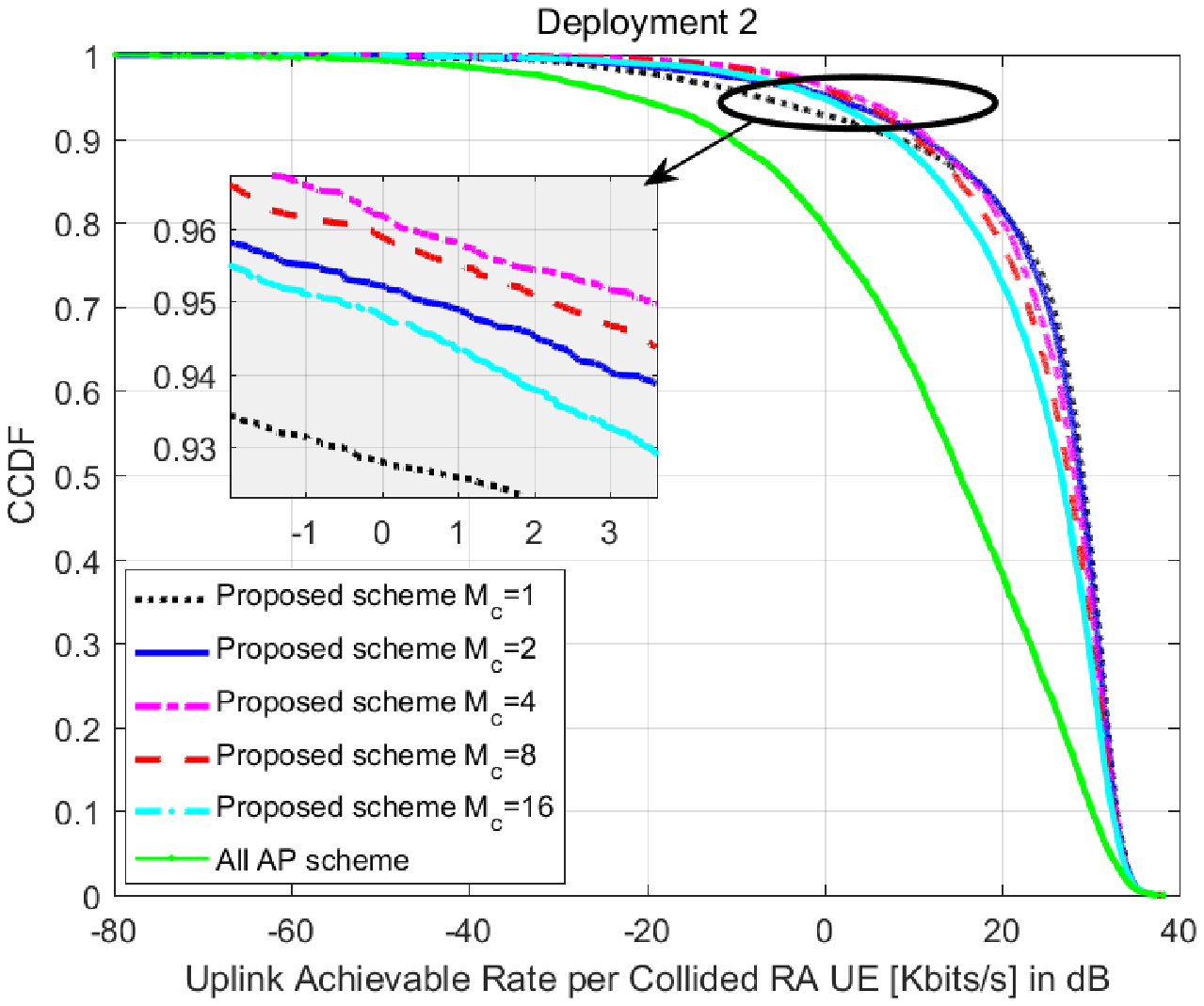}
\hspace{-4.1ex}
\includegraphics[width=2.3in]{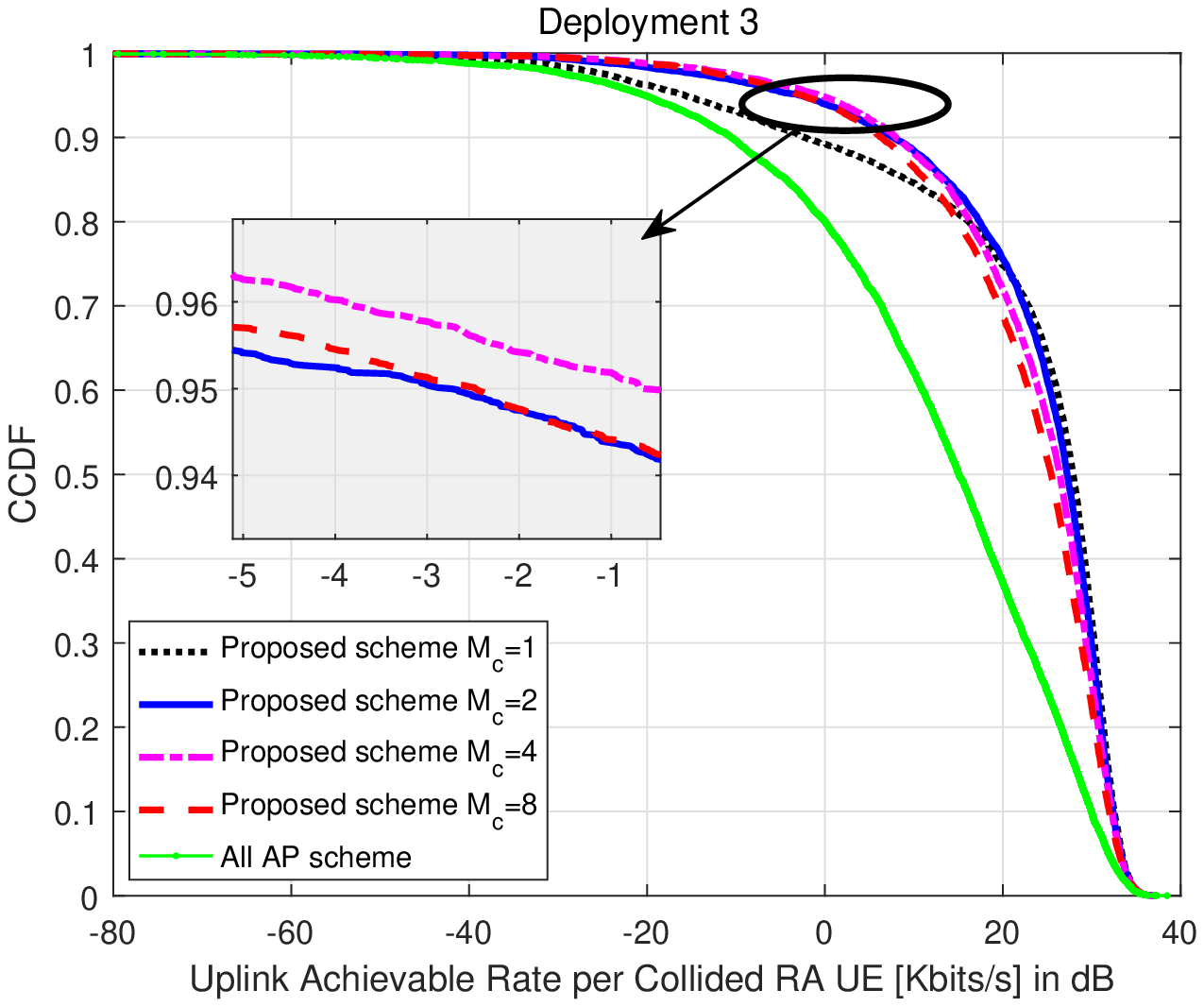}

\caption{ Uplink achievable rate CCDF per collided RA UE of the proposed scheme, with different $M_\mathrm{c}$ and $\sigma_{\mathrm{SF}}=8$ dB.}

\label{fig6_1}
\end{figure*}

We first investigate the performance of the proposed DNN based $K$-means AP clustering scheme in terms of $\mathcal{R}_{1}$ in GFRA with different $M_\mathrm{c}$ in the considered three deployments with $\sigma_{\mathrm{SF}}=8$. In Figure \ref{fig6_1}, uplink achievable rate complementary cumulative distribution functions (CCDF) per collided RA UE with different $M_\mathrm{c}$ are presented. The performance of the all-AP scheme is also considered in the figure as a baseline. As observed, compared to the all-AP scheme, the proposed DNN based scheme is able to significantly improve the achievable rate for collided RA UEs with a wide range of $M_\mathrm{c}$, which verify its effectiveness in resolving the preamble collision in GFRA. In addition, different $M_\mathrm{c}$ leads to different performance. Evidently, neither a too small
nor a too large value of $M_\mathrm{c}$ is desirable in the proposed scheme. By comparing the $95$\%-likely performance, it is clear to see that, throughout the three deployments, the proposed DNN based scheme with $M_\mathrm{c}=4$ provides the highest achievable rate, for example, achieving about $16$ dB performance gain to the case of $M_\mathrm{c}=1$ and $4$dB to the case of $M_\mathrm{c}=16$ in deployment $1$. These
observations validate the analysis provided at the end of Section IV. In the following, we further study the performance of the proposed DNN based scheme with only $M_\mathrm{c}=4$ in different deployments.

%with $M_\mathrm{c}=4$ as a representative example

\begin{figure*}[!h]
\centering

\includegraphics[width=2.3in]{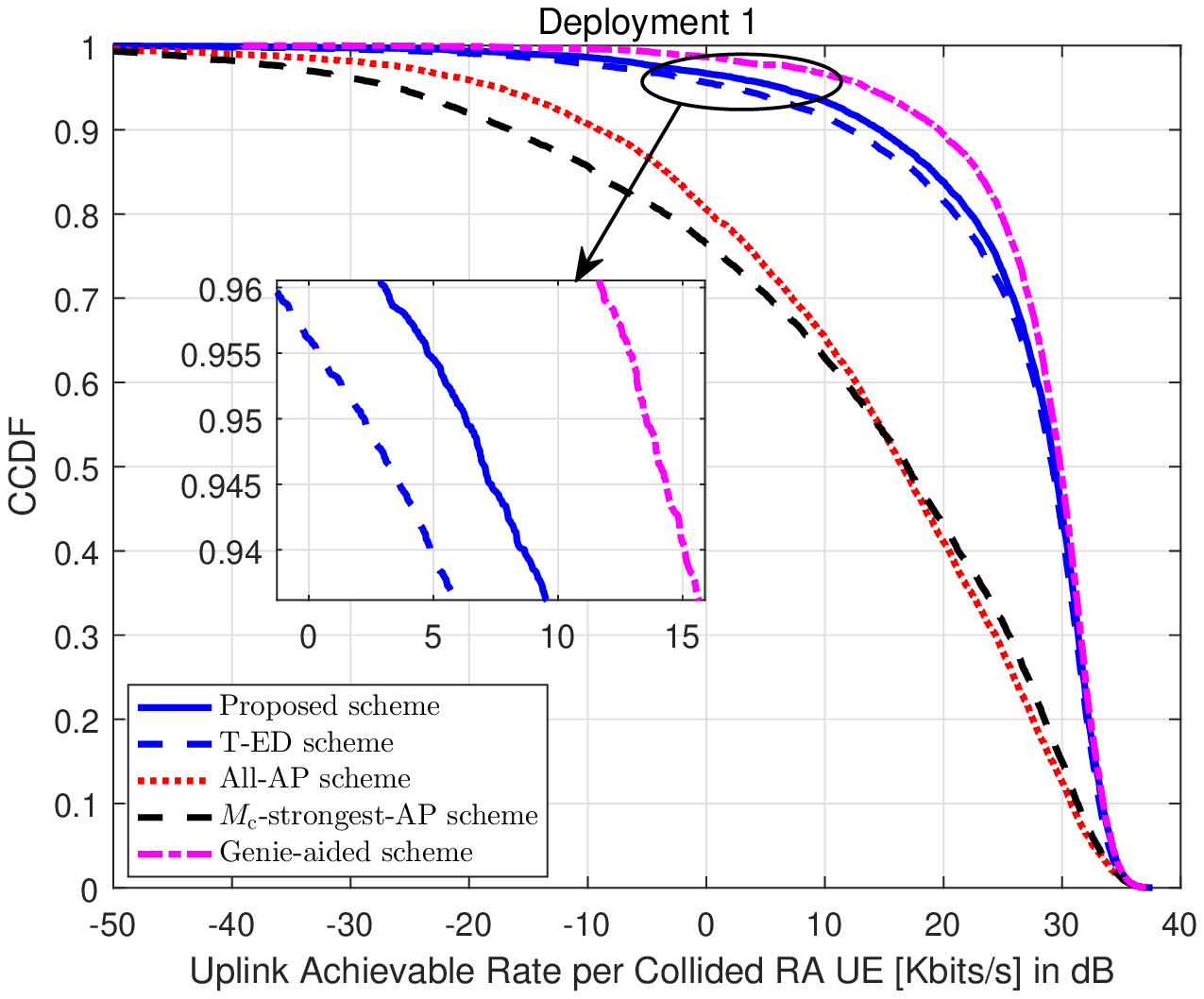}
\hspace{-4.1ex}
\includegraphics[width=2.3in]{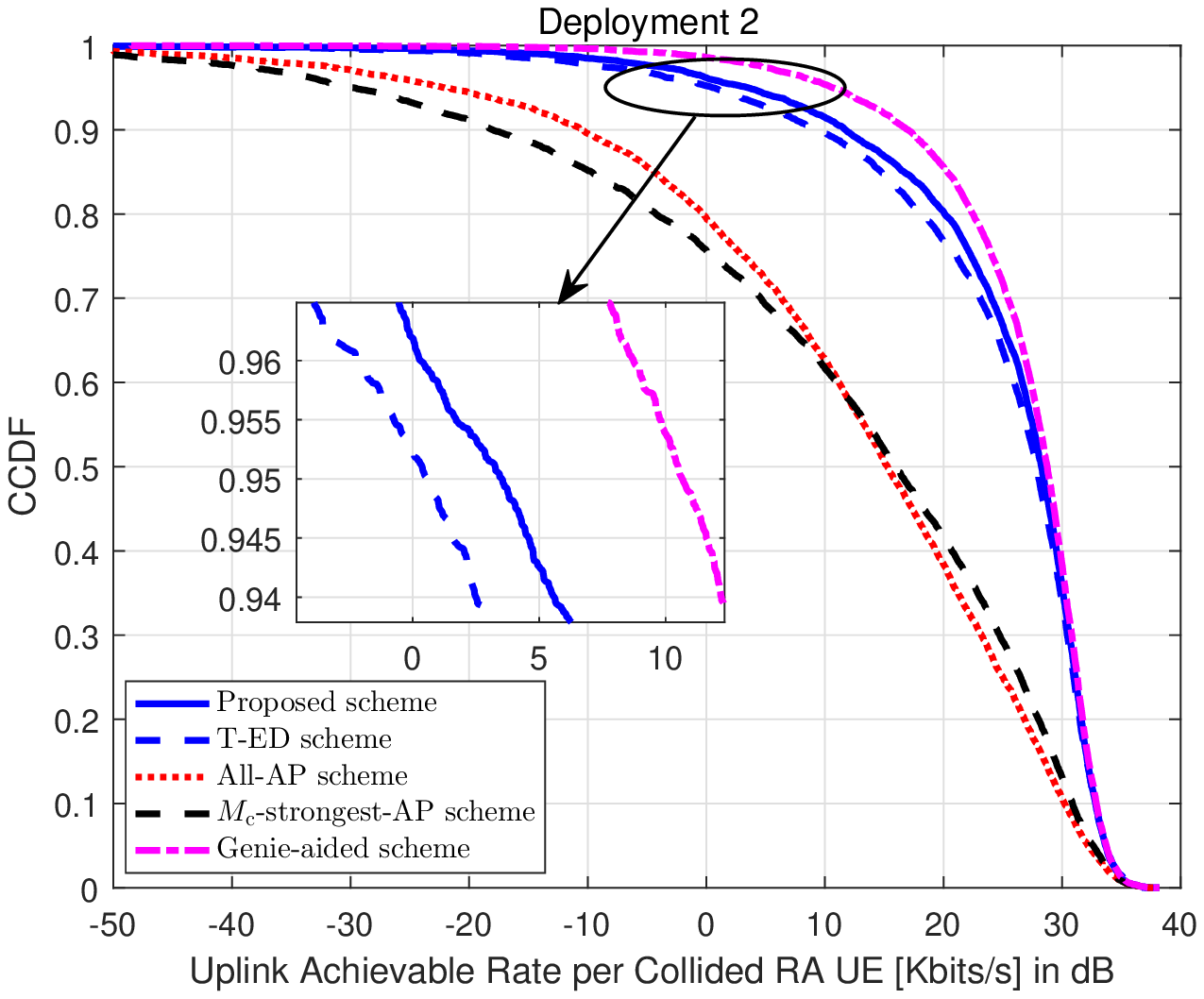}
\hspace{-4.1ex}
\includegraphics[width=2.3in]{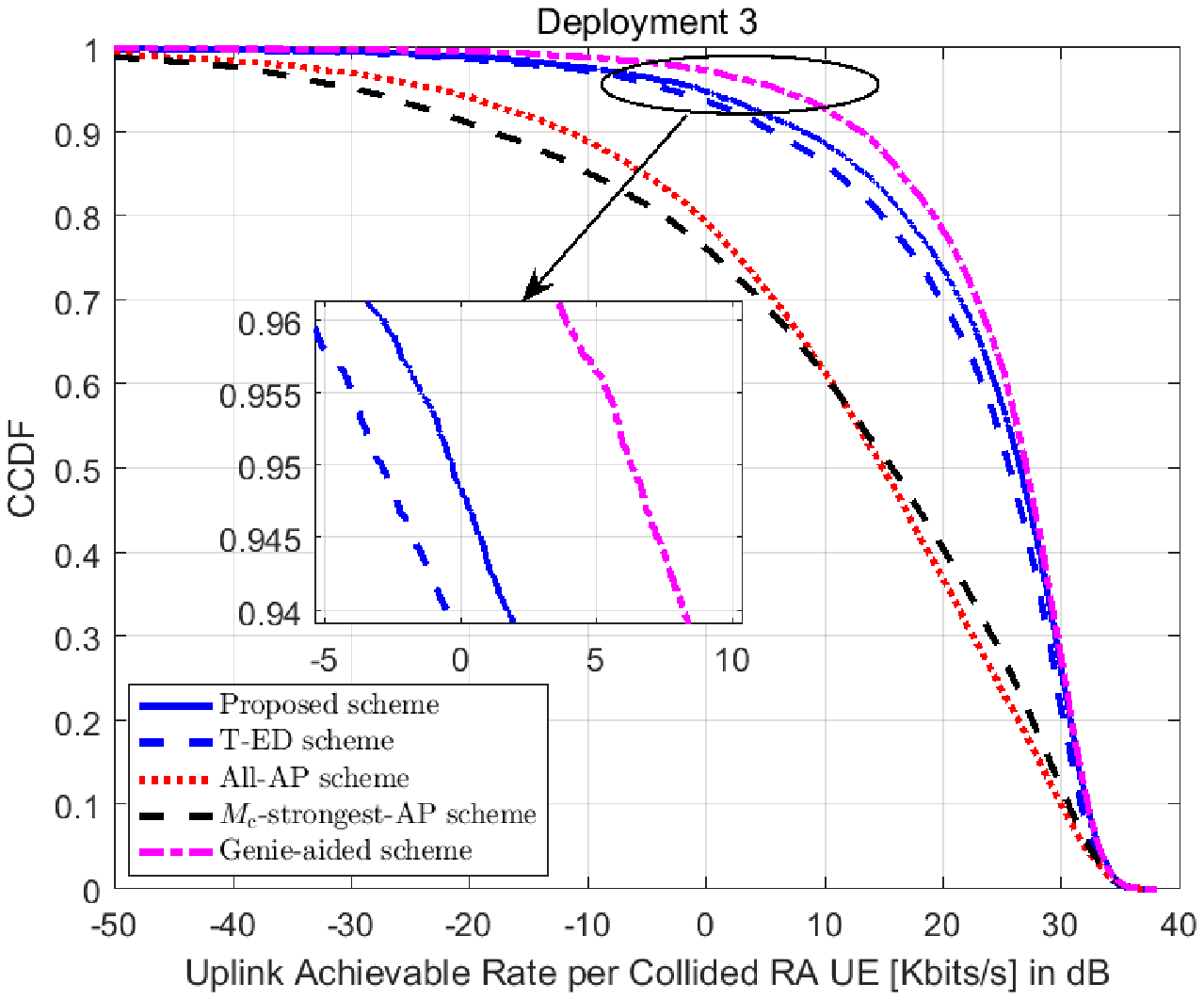}

\caption{ Comparisons of uplink achievable rate CCDF per collided RA UE among different schemes, with $\sigma_{\mathrm{SF}}=8$ dB and $M_\mathrm{c}=4$.}

\label{fig6}
\end{figure*}

\begin{figure}[!h]
	%\vspace{-0.1cm}  %调整图片与上文的垂直距离
	%%\setlength{\abovecaptionskip}{-0.1cm}   % 调整图片标题与图距离
	%\setlength{\belowcaptionskip}{-0.2cm}   %调整图片标题与下文距离
	\centering
	\includegraphics[width=4in]{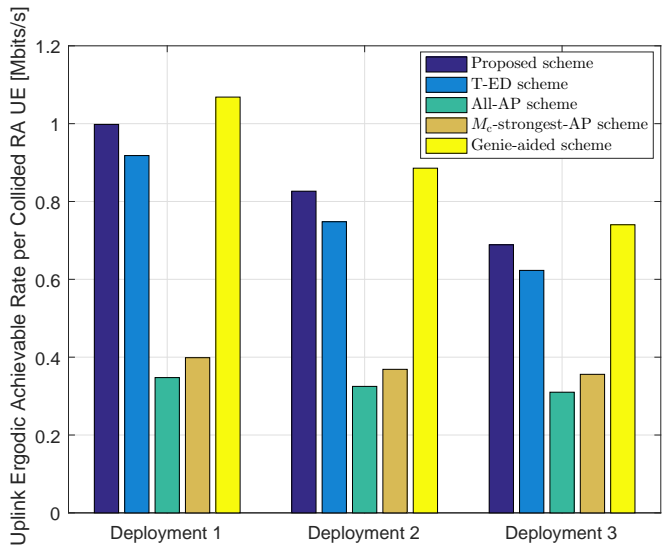}
	% where an .eps filename suffix will be assumed under latex,
	% and a .pdf suffix will be assumed for pdflatex; or what has been declared
	% via \DeclareGraphicsExtensions.
	%\captionsetup{justification=centering}
	\caption{Comparisons of uplink ergodic achievable rate per collided RA UE among different schemes, with $\sigma_{\mathrm{SF}}=8$ dB and $M_\mathrm{c}=4$.} \label{fig7}
\end{figure}

In Figure \ref{fig6}, the uplink achievable rate CCDF per collided RA UE in different deployments are shown.
Evidently, the proposed DNN and T-ED based schemes both perform much better than the all-AP scheme, which validates the effectiveness of the proposed framework that incorporates preamble multiplicity estimation and AP clustering to address the preamble collision problem in distributed mMIMO.
Since the DNN can provide more accurate preamble multiplicity estimation than the T-ED (as shown in the example given in Section \ref{T-ED}), the DNN based scheme outperforms the T-ED based scheme under different deployments. For instance, around $4$ dB performance gain in term of the $95$\%-likely uplink achievable rate per collided RA UE can be achieved in deployment $1$, which is approximately equivalent to $150$\% performance improvement.
Besides, both the all-AP and $M_\mathrm{c}$-strongest-AP schemes provide poor $95$\%-likely achievable rate performance for collided RA UEs because these two schemes have no capability of dealing with preamble collision and as a result, introduce strong interference originating from the preamble collision. In addition, it is shown that both schemes in deployment $1$ only exhibit slightly better performance than that in deployment $2$. This is due to the fact that employing $S=2$ not only brings in an array gain to the desired signals, but also to the interference due to the preamble collision. Therefore, increasing $S$ has trivial benefit of improving the performance of these two schemes. Interestingly, we see that the all-AP scheme is more preferable to the $M_\mathrm{c}$-strongest-AP scheme in terms of the $95$\%-likely performance. This results from that, without preamble multiplicity information and sensible AP clustering, the $M_\mathrm{c}$-strongest-AP scheme leads to severe clustering errors, i.e., the $M_\mathrm{c}$ APs could be the neighboring APs of other collided RA UEs that are far away from the target RA UE, which make these APs contain little desired signals of the target RA UE, but strong interference from other collided RA UEs in most times. This implies the necessity of preamble multiplicity estimation to enable correct AP clustering.
Contrastingly, based on the proposed DNN based preamble multiplicity estimation, the proposed AP clustering scheme is significantly superior to the all-AP and $M_\mathrm{c}$-strongest-AP schemes. For instance in deployment $1$, compared to the two schemes, the $95$\%-likely achievable rate of a collided RA UE can be largely enhanced by $26$ dB and $34$ dB, respectively. This performance superiority is also validated in Figure \ref{fig7}, where the uplink ergodic achievable rates per collided RA UE in GFRA in different schemes and deployments are presented. Compared to the all-AP and $M_\mathrm{c}$-strongest-AP schemes, the proposed DNN based scheme is able to provide an improvement of $187$\% and $150$\%, respectively, in terms of the ergodic rate per collided RA UE in deployment $1$. Similar improvement can also be seen in other two deployments.
Furthermore, it is shown that the CCDF curves of the proposed DNN based scheme are close to those of the genie-aided scheme. In deployment $1$, for example, a $7$ dB performance gap is observed in terms of $95$\%-likely performance. Considering that the performance of the proposed DNN based scheme is achieved without any CSI information of RA UEs, we claim that this performance gap is quite acceptable.
This insight is also supported by the results in Figure \ref{fig7}. In particular, only around $7.0$\% and $8.5$\% ergodic rate losses in deployment $1$ and deployment $3$ are observed respectively, by comparing the proposed DNN based scheme to the genie-aided one. %without need of knowing location and CSI information of the RA UEs.
These observations indicate that the proposed machine learning based framework solution is able to achieve a near-optimal performance under preamble collision, validating its effectiveness in terms of preamble collision resolution under given deployments of distributed mMIMO.

\section{Conclusion}
In this paper, a novel machine learning based framework was proposed to mitigate the impact of preamble collision on the performance of collided RA UEs in GFRA with distributed mMIMO.
By taking advantages of signal spatial sparsity in distributed mMIMO and sporadic traffic pattern of mMTC RA UEs, we first developed a tailored DNN to enable the preamble detection and estimate the preamble multiplicity in GFRA, where a T-ED method was also proposed for performance comparison. Under practical wireless environments and different deployments of distributed mMIMO, we analyzed and compared the performance of the proposed DNN and showed that decent estimation accuracy and reliability can be achieved. With the estimated preamble multiplicity, we then proposed a $K$-means AP clustering algorithm to cluster the neighboring APs of collided RA UEs rather than all the APs to decode the received data individually. Simulation results verified the effectiveness of the proposed schemes in preamble collision resolution in GFRA. Particularly, as examples shown in the simulation, the proposed DNN based scheme is able to achieve a close performance to the genie-aided scheme in terms of uplink achievable rate per RA UE under preamble collision. In the considered deployments of distributed mMIMO, the proposed DNN based scheme provided its best performance when $M_\mathrm{c}=4$, which exhibits a significant performance gain of up to $26$ dB over the all-AP scheme. %without preamble multiplicity detection and AP clustering.

The machine learning based framework solution could be served as a groundwork for preamble collision resolution in GFRA with distributed mMIMO. To further improve the performance, other powerful and efficient machine learning enabled solutions can be explored. For example, it might be better to integrate the preamble multiplicity estimation and AP clustering into one machine learning based algorithm, which is left for future research.

%We first use a toy example in distributed mMIMO to theoretically show that the performance impairment of each collided RA UE in GFRA could be potentially eliminated by only employing its neighbouring APs to serve instead of all the APs. Motivated by this,

%and traffic sporadicity

% trigger a \newpage just before the given reference
% number - used to balance the columns on the last page
% adjust value as needed - may need to be readjusted if
% the document is modified later
%\IEEEtriggeratref{8}
% The "triggered" command can be changed if desired:
%\IEEEtriggercmd{\enlargethispage{-5in}}

% references section

% can use a bibliography generated by BibTeX as a .bbl file
% BibTeX documentation can be easily obtained at:
% http://www.ctan.org/tex-archive/biblio/bibtex/contrib/doc/
% The IEEEtran BibTeX style support page is at:
% http://www.michaelshell.org/tex/ieeetran/bibtex/
%\bibliographystyle{IEEEtran}
% argument is your BibTeX string definitions and bibliography database(s)
%\bibliography{IEEEabrv,../bib/paper}
%
% <OR> manually copy in the resultant .bbl file
% set second argument of \begin to the number of references
% (used to reserve space for the reference number labels box)
\normalem
% that's all folks
\end{document}